\documentclass[preprint,prd,amssymb,floatfix,nofootinbib,balancelastpage,superscriptaddress,amsmath,floats]{revtex4}

\pdfoutput=1

\usepackage{epsfig}
\usepackage{latexsym}
\usepackage{graphicx}
\usepackage{amssymb}
\usepackage{subfigure}
\def\bea{\begin{eqnarray}}
\def\eea{\end{eqnarray}}
\def\be{\begin{equation}}
\def\ee{\end{equation}}

\begin{document}

\DeclareGraphicsExtensions{.pdf,.png,.gif,.jpg}

\title{Can we discover dual-component thermal WIMP dark matter?}

\author{Stefano Profumo}
\email{profumo@scipp.ucsc.edu}
\affiliation{%
Santa Cruz Institute for Particle Physics and Department of Physics,\\ University of California, Santa Cruz CA 95064
}

\author{Kris Sigurdson}
\email{krs@physics.ubc.ca}
\affiliation{Department of Physics and Astronomy, University of British Columbia, Vancouver, BC V6T 1Z1, Canada}

\author{Lorenzo Ubaldi}
\email{ubaldi@physics.ucsc.edu}
\affiliation{%
Santa Cruz Institute for Particle Physics and Department of Physics,\\ University of California, Santa Cruz CA 95064
}


\begin{abstract}
We address the question of whether the upcoming generation of dark matter search experiments and colliders will be able to discover if the dark matter in the Universe has two components of weakly interacting massive particles (WIMPs). We outline a model-independent approach, and we study the specific cases of (1) direct detection with low-background 1 ton noble-gas detectors and (2) a 0.5 TeV center of mass energy electron-positron linear collider. We also analyze the case of indirect detection via two gamma-ray lines, which would provide a verification of such a discovery, although multiple gamma-ray lines can in principle originate from the annihilation of a single dark matter particle. For each search ``channel'', we outline a few assumptions to relate the very small set of parameters we consider (defining the masses of the two WIMPs and their relative abundance in the overall dark matter density) with the relevant detection rates. We then draw general conclusions on which corners of a generic dual-component dark matter scenario can be explored with current and next generation experiments. We find that in all channels the ideal setup is one where the relative mass splitting between the two WIMP species is of order 1, and where the two dark matter components contribute in a ratio close to 1:1 to the overall dark matter content of the Universe.  Interestingly, in the case of direct detection, future experiments might detect multiple states even if only $\sim$10\% of the energy-density of dark matter in the Universe is in the subdominant species.
\end{abstract}





\maketitle

\section{Introduction}\label{sec:intro}

The microscopic nature of the dominant component of the matter density of the Universe, the non-baryonic dark matter, is presently unknown. While several theoretically compelling hypotheses have been put forward to explain the fundamental physics of particle dark matter, the paradigm of weakly interacting massive particles  (WIMPs), has increasingly attracted more attention and has been closely scrutinized both from the theoretical and from the experimental standpoint (for comprehensive reviews of WIMP dark matter candidates and their detection, see, e.g., \cite{Jungman:1995df,Bergstrom:2000pn,Bertone:2004pz,Hooper:2007qk}; for a pedagogical review of dark matter astrophysics see, e.g., \cite{dmastro}). 

 Remarkably, the standard calculation of the  thermal freeze-out of species in the early Universe \cite{standardcalc} predicts that  WIMPs generically have a relic density today which is very close to the measured density of dark matter $\Omega_{\rm DM}h^2 \simeq 0.11$ \cite{wmap}. While elusive, WIMPs are in principle detectable since in standard scenarios they can interact non-gravitationally with ordinary matter via weak interactions. Two approaches have been pursued to detect non-graviatational signals from WIMPs in our Galaxy: {\em direct detection} and {\em indirect detection}.  Direct detection hinges upon  directly observing the scattering of WIMPs off the nuclei of ordinary matter in very low background terrestrial environments.  Indirect detection, instead, aims to detect and identify the annihilation debris of WIMPs which annihilate in the Galaxy, including antimatter (positrons, antiprotons, antideuterons), high-energy neutrinos, and gamma rays. Last but not least, WIMPs are predicted to exist in several potential extensions to the standard model of particle physics including, for example, supersymmetry, models with universal extra dimensions or little Higgs models.  The new physics expected in these models can manifest itself as distinctive signatures at high-energy colliders and WIMPs, in many cases, are the final result of (potentially long) decay chains when new particle states are pair-produced in high-energy collisions. A typical signature at colliders includes a large missing transverse momentum associated with the WIMP escaping the detector, and a characteristic series of energetic jets and leptons streaming off of the WIMP-generating decay chain.

When exploring the realm of unknown physics it is certainly a reasonable approach to invoke Occam's razor and to assume as a working hypothesis the simplest scenario consistent with observations.
Given the persuasive theoretical and observational indications discussed above, if we restrict our attention to a WIMP dark matter scenario, it makes sense to postulate that a single stable WIMP species accounts for all of the dark matter. Yet, while a perfectly acceptable and reasonable assumption, this might be an oversimplification of the truth. The dark-matter sector might not be as simple as to be made up by one single component, and assuming so may obscure new physics in the dark sector.  We believe it is legitimate to entertain the hypothesis that dark-sector physics reflects the variety and complexity of elementary constituents that populate the ordinary-matter sector, and it is this hypothesis we scrutinize in this work.

Scenarios with more than one dark matter particle constituent have been
envisioned and investigated in the past (see our review and discussion  
in
sec.~\ref{sec:related}). 
In this paper we focus on the possibility that the dark matter has  
two WIMP
constituents. We have two goals:
\begin{itemize}
\item we want to establish whether current experiments that focus on WIMP
detection have the capability to actually discriminate more than one  
WIMP species;
\item we want to answer the phenomenological question:   
if signatures
of more than one WIMP species can be detected, which portions of the
WIMP parameter space are favored?
\end{itemize}
In the analysis we keep in
mind several theoretical principles that could guide model building of a
multiple WIMP scenario, where the WIMPs have similar physical properties.

On more quantitative grounds, the scope of the present study is twofold.  Firstly, we aim to identify which multi-particle dark matter models might be discovered with present and next generation dark matter detection and collider experiments; secondly, we aim to compare and cross-correlate --- using appropriate assumptions to relate the relevant physical parameters --- the potential of direct, indirect and collider searches for multi-component particle dark matter. We must stress that, whereas direct detection and collider searches could lead to the discovery of a dual-component scenario, indirect detection could not provide a discovery in its own right. This has to do with a fundamental problem related to the detection of multiple monochromatic lines: given two lines, one cannot distinguish a second line arising from a second WIMP from an additional line due to other annihilation channels of the first WIMP (annihilations into $Z\gamma$ or $H\gamma$ for example). But if the discovery is made through the other search channels (direct detection and collider), then indirect detection via two gamma-ray lines could provide a valuable verification. Keeping in mind this caveat, in this paper we also analyze the indirect detection case in detail, in order to compare to the direct detection and collider results. 
  
The manuscript is organized as follows: In Section~\ref{sec:model} we describe the physical parameters of our model and survey our main results. Section~\ref{sec:direct} focuses on direct dark matter detection, Section~\ref{sec:colliders} on collider searches, Section~\ref{sec:indirect} on indirect detection. In Section~\ref{sec:related} we comment on related work and potential multi-particle dark matter models, and in Section~\ref{sec:conclusions} we conclude with a summary.

\section{Model setup and summary of results}\label{sec:model}

We consider a scenario where two particle species $\chi_1$ and $\chi_2$ are stable on cosmological scales, and contribute to the total dark matter energy density $\Omega_{\rm DM}$ with energy densities, respectively, $\Omega_{\chi_1}$ and  $\Omega_{\chi_2}$. We assume that the whole of the non-baryonic dark matter is made up of $\chi_1$ and $\chi_2$ particles, {\em i.e.} $\Omega_{\rm DM}=\Omega_{\chi_1}+\Omega_{\chi_2}$, and that both particles are weakly interacting massive particles. We define the ratio of relic abundances of the two species as
\be \label{eq:w}
w\equiv\frac{\Omega_{\chi_1}}{\Omega_{\chi_2}}\, ,
\ee
which allows us write
\be
\Omega_{\chi_1}=\frac{w}{1+w}\Omega_{\rm DM}\quad {\rm and}\quad \Omega_{\chi_2}=\frac{1}{1+w}\Omega_{\rm DM}.
\ee
As long as $\langle\sigma_{\chi_{1,2}\chi_{1,2}}v\rangle_{\rm tot}$ is dominated by its $s$-wave amplitude we have the well-known relation \cite{standardcalc}:
\be
\Omega_{\chi_{1,2}}\propto\frac{1}{\langle\sigma_{\chi_{1,2}\chi_{1,2}}v\rangle_{\rm tot} (T=0)}\, ,
\ee
which is independent of mass up to logarithmic corrections.  The parameter $w$ is thus also typically equal to to the ratio of total annihilation cross sections $r$ as
 \be
r \equiv \frac{{\langle\sigma_{\chi_{2}\chi_{2}}v\rangle_{\rm tot} (T=0)}} {{\langle\sigma_{\chi_{1}\chi_{1}}v\rangle_{\rm tot} (T=0)}}
\simeq w.
\label{eq:wcross}
\ee

For most of the following discussion we take particle 1 to be lighter than particle 2, {\em i.e.} $m_2>m_1$ and we define a parameter $\Delta$ corresponding to the relative mass splitting
\be
\Delta \equiv\frac{m_2-m_1}{m_1}.
\ee
In some cases we will trade $\Delta$ for $m_2$, relax the assumption $m_2>m_1$, and plot $m_2$ vs $m_1$.

Given this setup, we wish to address the question of the {\em detectability} of at least one of the dark matter species, and the question of our ability to {\em discriminate} between a single and a dual-component dark matter scenario, in the case that both species are detectable. In order to proceed, we shall need to outline sets of minimal, further assumptions, pertinent to various dark matter detection techniques, as well as to collider searches. In particular, in this study we will address: 
\begin{itemize}
\item[(I)] the {\em direct detection} of dark matter, proceeding via the detection of recoil energy resulting from the elastic scattering of WIMPs off of nucleons (for a review see e.g. \cite{Munoz:2003gx}); 
\item[(II)] {\em collider searches} for new physics, featuring the end-production of two stable, weakly interacting massive particles that escape detection carrying away significant missing energy and momentum, with a high center of mass energy $e^+e^-$ linear collider. For simplicity, we do not consider here the otherwise extremely interesting phenomenology of scenarios with two stable weakly interacting states that can escape detection at the LHC, since discussing in any detail that scenario would require a number of highly model-dependent assumptions (including the detailed mass spectrum of particles that participate in the cascade decays of the produced primary particle pairs, branching fractions, decay modes, etc.);
\item[(III)] the {\em indirect detection} of dark matter, specifically via the (likely) cleanest possible signature of WIMP pair annihilation, namely the detection of a monochromatic gamma-ray line from WIMP pair annihilation into two photons (with energies equal to the WIMP mass) with the Fermi Large Area Telescope (LAT) \cite{Atwood:2009ez}.
\end{itemize}
We define below the additional parameters needed, within the minimal and model-independent particle setup we utilize here, to quantitatively discuss the three experimental probes of dark-matter physics listed above.

\subsection{Direct Detection}

\noindent The rate at which WIMPs elastically transfer energy to nuclei depends on astrophysical parameters, as the local density and velocity distribution of WIMPs, and on particle physics parameters, as the WIMP mass and its scattering cross section off protons and neutrons. We assume here, for simplicity, that the scattering cross section of either dark matter particle species is identical for both neutrons and protons ($\sigma_{\chi_{1,2}n}$), and consider the following parameter space:
\be \label{eq:dirpar}
{\rm\bf Direct\ detection\ Parameter\ Space}:\quad m_1,\ \Delta \ {\rm or} \ m_2,\ w,\ \sigma_{\chi_{1}n}, \ \sigma_{\chi_{2}n}.
\ee
Given that $m_1$ and $\Delta$ specify the WIMPs masses, and $w$ the relative abundance (which we assume to be the same locally and on cosmological scales), the parameter space above fully specifies the WIMP properties relevant to calculate the direct detection rates.

In particular cases, it can be useful to postulate a relationship between the WIMPs pair annihilation cross section and the scattering cross section of WIMPs off nuclei, and between the scattering cross sections of the two WIMP species under consideration here. As a guideline,  one can assume that the particles exchanged in WIMP-nucleon scattering are the same ones that contribute (in a crossed-channel) to the WIMP pair annihilation in the Early Universe. With the case of supersymmetry in mind, if we assume this exchanged particle is a ``squark'' $i=1,2$, and that the squark mass $m_{\widetilde S}$ is much larger than the DM mass, we can write:
\be
\sigma_{\chi_i n}\sim \frac{m_n^4}{m_W^2\ m_{\widetilde S_i}^4}
\ee
and
\be
\sigma_{\chi_i\chi_i}\sim \frac{m_{\chi_i}^2}{\ m_{\widetilde S_i}^4}.
\ee
One therefore can estimate
\be \label{eq:wsigmasdir}
r_{\chi n} \equiv \frac{\sigma_{\chi_{2}n}}{\sigma_{\chi_{1}n}} \simeq  \frac{m_{\chi_1}^2}{m_{\chi_2}^2}\frac{\langle\sigma_{\chi_{2}\chi_{2}}v\rangle}{\langle\sigma_{\chi_{1}\chi_{1}}v\rangle}\simeq \frac{m_{\chi_1}^2}{m_{\chi_2}^2}w.
\ee

When appropriate and explicitly specified we will assume the above relation to hold and use it to relate one scattering cross section to the other according to Eq.~(\ref{eq:wsigmasdir}), e.g.: $\sigma_{\chi_2 n}\simeq w\sigma_{\chi_1 n}(m_{\chi_1}/m_{\chi_2})^2$.

\subsection{Collider Searches}
\label{subsec:collider}

\noindent Outlining the phenomenology of models with two stable massive weakly interacting particles that end the decay chain of new particle species produced in high-energy collisions is a formidable and complex task. Even simply counting event rates that involve one or the other particle involves specifying a number of features of the particle model at hand, including the production cross section for the heavy particles that subsequently decay into the stable WIMPs and the topology and branching ratios for the decay chains that lead to the escaping WIMP(s). In addition, in the present analysis we aim at understanding whether it is possible to not only detect the existence of WIMPs, but to {\em discover} more than one stable WIMP species and to {\em discriminate} this scenario from one where there is only a single stable WIMP. For a model-independent analysis, in the case of colliders, an unambiguous experimental handle on the particle mass is likely needed. Although several studies argue that, for specific models or event topologies the Large Hadron Collider (LHC) will provide such handle in single WIMP models (see e.g. \cite{baertatabook} for extensive discussions in the case of supersymmetric models, and Ref.~\cite{Cheng:2008mg} for specific event topologies), it is obvious that a hadron collider is not the ideal environment for precision studies of the absolute mass scale of new particles. The main reason is simply that the center-of-mass energy of the underlying event in a hadron collider is not known. We leave the question of model-specific scenarios with multiple stable WIMPs at the LHC for future dedicated studies.

We instead articulate in the present study an analysis of a high-energy electron-positron collider that might follow the LHC at some point in the future. With such a machine it will in general be possible to tune the center of mass energy of collisions, and to have an extremely accurate account of the mass scale of new particles. In this case, the ability to statistically detect a signal and to discriminate between a single and multiple stable WIMP scenarios (where stable naturally alludes to collider time-scales) is a well defined question that depends only on the number of events produced in high-energy collisions and assumptions about the experimental features of the collider detectors and relevant backgrounds. We give details on our assumptions about all these parameters in Sec.~\ref{sec:colliders}, where we consider, as a prototypical future electron-positron collider, the so-called International Linear Collider (ILC).

\begin{figure}[!ht]
  \subfigure[]{
     \label{fig:pp}
    \includegraphics[width=60mm]{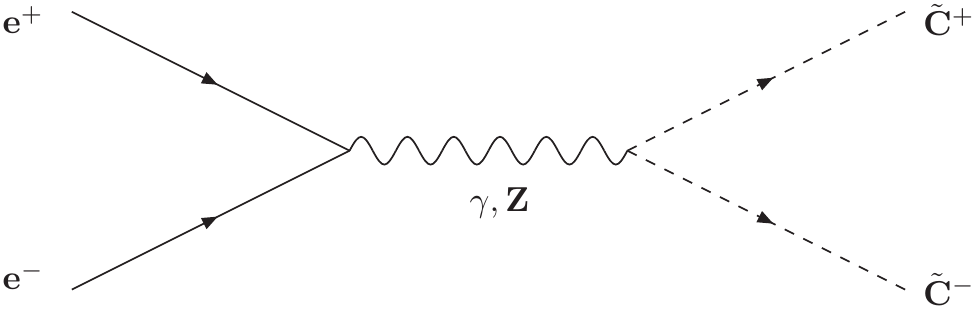}}
  \hspace{5mm}
  \subfigure[]{
    \label{fig:an}
    \includegraphics[width=30mm]{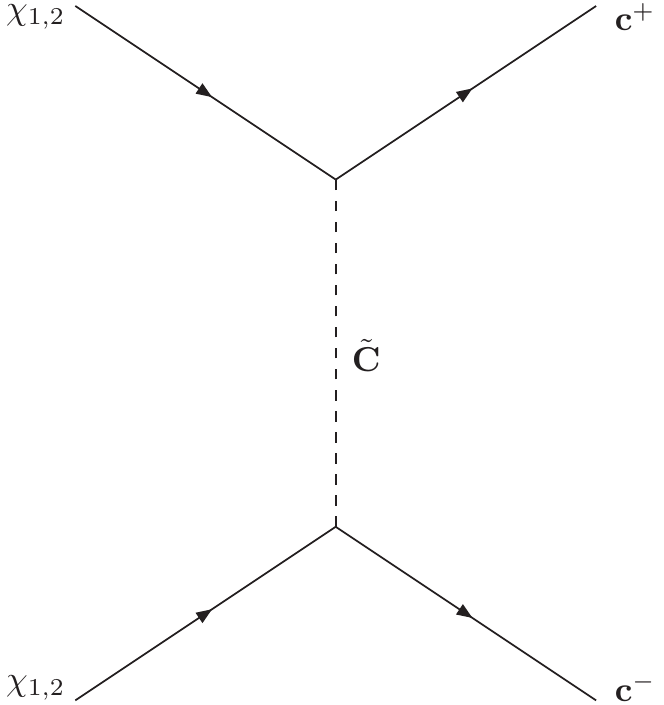}}
  \hspace{5mm}
  \subfigure[]{
    \label{fig:decay}
    \includegraphics[width=30mm]{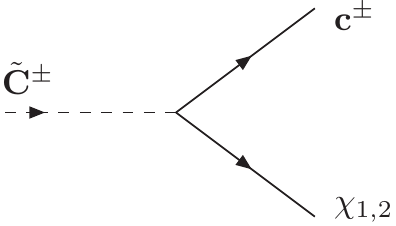}}
  \caption{\textit{Feynman diagrams.} \ref{fig:pp} Pair production of $\tilde{C}^\pm$ at collider, \ref{fig:an} Dark matter annihilation, \ref{fig:decay} Decay of $\tilde{C}^\pm$.}
  \label{fig:diagrams}
\end{figure}

While we give full details of our assumptions in Sec.~\ref{sec:colliders}, we wish to first briefly outline here the scenario and parameter space we shall consider.  
As it is unlikely that stable WIMPs will be the most easily produced particles at the collider, we envision a simple scenario where a pair of heavy charged particles $\widetilde C$ of mass $m_{\widetilde C}$ are produced that subsequently decay into either $\chi_1$ or $\chi_2$ particles.  In the limit that processes involving the $\widetilde C$ particle dominate the total pair annihilation cross section (for instance via $t-$ or $u-$channel annihilation into standard model fermions, as is the case of neutralino pair -annihilation into fermion-antifermion final states via sfermion exchange) we have the relation
\be
r_{\alpha^2} \equiv \frac{\alpha_2^2}{\alpha_1^2} \simeq \frac{\langle\sigma_{\chi_{2}\chi_{2}}v\rangle}{\langle\sigma_{\chi_{1}\chi_{1}}v\rangle} \simeq w
\label{eq:g2ratio}
\ee
between the ratio of the squared structure constants $r_{\alpha^2}$ and the ratio of relic abundances $w$. The relevant Feynman diagrams are shown in Figure~\ref{fig:diagrams}.
Here $\alpha_{1,2} = \frac{g_{1,2}^2}{4\pi}$, $g_{1,2}$ is the coupling constant for the vertex involving ${\widetilde C^{\pm}}, \ \chi_{1,2}, \  {c}^{\pm}$, where $c^{\pm}$ denotes a charged standard model particle.  Using Eq.~(\ref{eq:g2ratio}), we can then relate the branching ratios for $\widetilde C$ decay to $\chi_{1,2}$ to $w$.
Accounting for the two WIMP masses (determined by $m_1$ and the relative mass splitting $\Delta$), the simplest parameter space we can envision for this scenario is:

\be \label{eq:collpar}
{\rm\bf Linear\ Collider\ Parameter\ Space}: \quad m_1, \ \Delta \ {\rm or} \ m_2,\ r_{\alpha^2} \simeq w, \ m_{\widetilde C}.
\ee

\subsection{Indirect Detection: Monoenergetic $\gamma$ Rays}

\noindent In the era of precision gamma-ray astronomy, a golden indirect detection channel is the search for monochromatic $\gamma$ rays resulting from the pair annihilation of two dark matter particles into two photons (for early studies see,  {\em e.g.}, Ref.~\cite{Bergstrom:1988fp}, see Ref.~\cite{Bergstrom:1997fh} for the complete calculation in the context of the minimal supersymmetric extension of the Standard Model (MSSM), and Ref.~\cite{Ferrer:2006hy} for the extension of that calculation to the next-to-MSSM). For non-relativistic (cold) dark matter particles, which pair-annihilate at very low relative velocities, the energy of the two resulting $\gamma$ rays equals the mass of the dark matter particle.

While in specific WIMP models the monochromatic gamma-ray line is a
suppressed annihilation mode, it is by all means a very striking  
signature
that could not be attributed to astrophysical processes. We might have
considered the gamma-ray continuum emission as well, which typically has
large fluxes. It is clear, however, that discriminating a dark matter
signature from more than one WIMP from astrophysical backgrounds (be it
from data taken from the galactic center or from the galactic halo  
diffuse
emission) will be a hopeless task for an instrument like Fermi. Since we
want to be here agnostic about other pair-annihilation channels, that
would be needed to compute the continuum gamma-ray emission, and in view
of the difficulties of background discrimination, we focus only on the
monochromatic line.

We can still not claim any discovery of a dual-component dark matter via detection of multiple lines, because of the complication discussed in the introduction, but this remains the cleanest signal that we could use as supporting evidence if such a scenario is found via other searches.

We could have also considered another smoking gun signature of dark  
matter
annihilation: a flux of high-energy neutrinos from the Sun or from the
center of the Earth. Again, though, this would imply the need for (a)
detailed assumptions on the WIMP annihilation modes, and thus would
introduce model-dependence, and (b) it is unclear if the spectral
information one gets from current or future neutrino telescopes would  
ever
allow for a robust discrimination of scenarios with more than one WIMP.
Though interesting, we thus choose not to entertain this indirect
detection channel in the present analysis.

Given that dark matter is {\em dark} (i.e. it has suppressed electro-magnetic interactions), the pair annihilation cross section of WIMPs into two monochromatic photons is, in generic WIMP models \cite{Jungman:1995df,Bergstrom:2000pn,Bertone:2004pz,Hooper:2007qk}, a loop-suppressed process. In specific constructions, however, the monochromatic photon pair-production process can be enhanced, as shown for instance in the case of inert Higgs dark matter scenarios \cite{Gustafsson:2007pc}, models with extended gauge sectors and higher dimensional operators \cite{Dudas:2009uq} and string-inspired scenarios \cite{Mambrini:2009ad}. In the present model-independent approach, we indicate the relevant annihilation cross sections with the symbols:
\be
\sigma(\chi_{1,2}\ \chi_{1,2}\to \gamma\ \gamma )\equiv\sigma_{\chi_{1,2}\chi_{1,2}\to\gamma\gamma}\, .
\ee
For simplicity we assume that the process where $\chi_1\chi_2\to\gamma\gamma$ is suppressed or forbidden and we neglect it (given this would give an additional line signature, this is a conservative assumption). To  obtain a more tractable parameter space we reduce the number of parameters and make the simplifying assumption that the total rate of pair-annihilation is proportional to the rate at which the WIMPs pair annihilate into two photons,
\be
\sigma_{\chi_{1,2}\chi_{1,2}\to\gamma\gamma} \propto \langle\sigma_{\chi_{1,2}\chi_{1,2}}v\rangle_{\rm tot} (T=0).
\ee
This assumption is valid in the context of several particle dark matter models: for instance, in models such as those with universal extra dimensions where the new physics sector lies at a quasi degenerate mass scale $m_{\rm KK}$, both $\langle\sigma_{\chi\chi}v\rangle_{\rm tot}$ and $\sigma_{\chi\chi\to\gamma\gamma}$ are proportional to $m^{-2}_{\rm KK}$ \cite{Hooper:2007qk}. Similarly, Ref.~\cite{kaplanbergs} showed that in the case of supersymmetric Higgsinos (another paradigmatic ``minimal dark matter'' particle model \cite{minimaldm}), the ratio of the two cross sections is constant, modulo logarithmic corrections in the ratio of the weak scale to the particle dark matter mass scale. The approximate result of \cite{kaplanbergs} was confirmed by the exact calculation of Ref.~\cite{ulliobergs}. In addition, several examples of model-independent constructions where the two-photon mode has a roughly constant branching ratio to the total annihilation cross section are given in Ref.~\cite{bergsnell}.

Using Eq.~(\ref{eq:wcross}), we can then argue that  the ratio of the annihilation cross sections to two photons is
\be \label{eq:wsigmas}
r_{\gamma\gamma} \equiv \frac{\sigma_{\chi_{2}\chi_{2}\to\gamma\gamma}}{\sigma_{\chi_{1}\chi_{1}\to\gamma\gamma}} \simeq w\, .
\ee

This set of assumptions is generically satisfied in most particle dark matter models, unless special circumstances occur, {\em e.g.} coannihilation processes, or accidental suppressions or enhancements of the radiative pair-annihilation channel into two photons. Once the mass of the WIMPs is specified, and the pair annihilation rate into two photons is given, the rate for the indirect detection channel under consideration here is fully specified from the particle physics standpoint. We therefore take the parameter space for this indirect detection channel to be:
\be \label{eq:indpar}
{\rm\bf Indirect\ detection\ Parameter\ Space}:\quad m_1,\ \Delta \ {\rm or} \ m_2,\ r_{\gamma\gamma}\simeq w,\ \langle\sigma_{\chi_{1}\chi_{1}\to\gamma\gamma}v\rangle.
\ee
Our ability to make predictions therefore only depends, once these parameters are known, upon specifying the dark matter density distribution in a given direction, the astrophysical background, and the specifications of the instrumental response and performance. We provide details on all these, and present our results on this parameter space in Sec.~\ref{sec:indirect}.

\subsection{Summary of the Results}\label{sec:resultssummary}

\begin{figure}[t]
  \centering  
  \mbox{\hspace*{-1.5cm}\includegraphics[width=190mm]{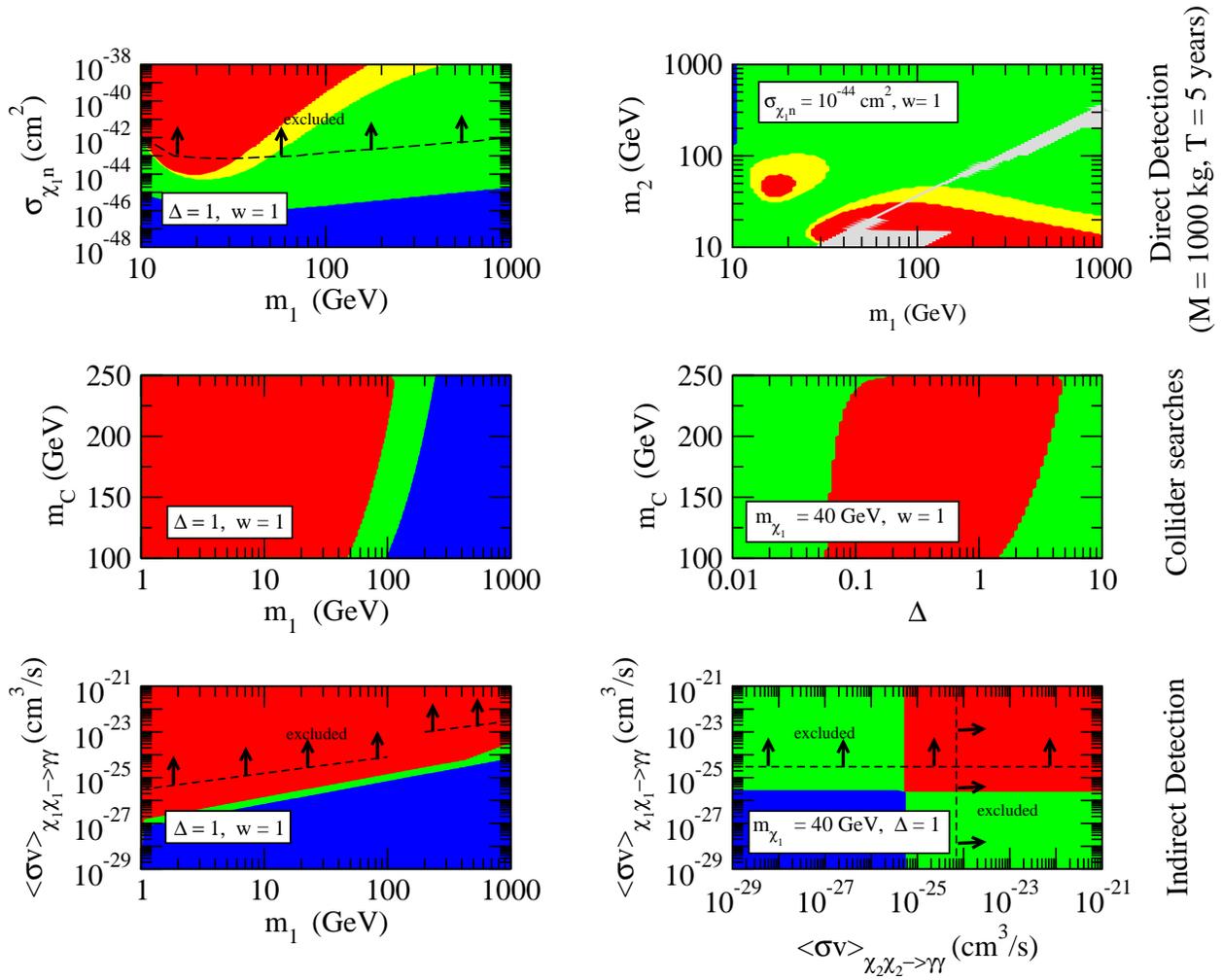}}\\[-1cm]
  \caption{\textit{Summary of our results}: A highlight of the results we present in this study for direct detection (upper panels), collider searches (central panels), indirect dark matter detection (lower panels). {\em Color coding}: blue stands for {\em no detection}, green for {\em detection without discrimination}, red and yellow for {\em detection with discrimination}.} 
  \label{fig:summary}
\end{figure}

For the ease of the reader we now briefly summarize our findings and outline the main highlights of our results in this subsection. Complete details on indirect detection, direct detection and collider searches are then given in the following three sections of the paper.  Figure~\ref{fig:summary} shows a selection of representative plots from the main body of the paper. We also take this opportunity to introduce the color-coding and conventions we shall use in the remainder of this study.

The top panels of Fig.~\ref{fig:summary} refer to direct detection (the results refer to a prototypical experiment employing 1 ton of Xenon and 5 years of data taking time), the middle panels to collider searches, the bottom panels to indirect dark matter searches. In all cases, the panels to the left refer to a WIMP setup where $\Delta=1$, hence the stable WIMP masses are in a 2:1 ratio ($m_2/m_1=2$), and where they contribute equally to the dark matter abundance $w=\Omega_{\chi_1}/\Omega_{\chi_2}=1$; the parameter space refer to the mass scale ($m_1=m_2/2$) for the x-axis, and to the variable that drives the event rate on the y-axis: respectively, the scattering rate off of nucleons for direct detection, the mass of the heavy species to be pair produced at a linear collider for collider searches, and the pair annihilation rate into two photons for indirect detection. The right panels inspect alternative slices of parameter space for each detection technique.

The color coding we employ here and in the rest of the paper is as follows:
\begin{itemize}
\item {\em Red} regions correspond to {\em detection with discrimination}: a combination of parameters such that we predict that the experiments should both detect a signal from both WIMP species, and discriminate the individual WIMPs (for instance, by distinguishing the two different WIMP masses in direct detection and collider searches).
\item {\em Green} regions correspond to {\em detection without discrimination}: in these regions of the parameter space, at least one of the WIMPs is conclusively discovered, but it is not possible to disentangle the presence of a second WIMP species. This is either because the second WIMP doesn't produce a large enough signal, or because the signal it produces cannot be experimentally distinguished from that coming from the first WIMP.
\item {\em Blue} regions indicate portions of the parameter space with {\em no detection}, i.e. where the experimental setups we consider would not see any statistically significant signal from either WIMP species.
\end{itemize}
In the case of direct detection, assumptions on the systematic control of background play an important role. We therefore also consider a

\begin{itemize}
\item  {\em Yellow} region, where one would have {\em detection with discrimination} (as in the red regions) in the optimistic limit of {\em perfect background control}, i.e. if one assumes that the rate of background events is perfectly known. 
\end{itemize}
This clearly extends the red region to wider regions of parameter space, where an imperfect background knowledge would impair the experimental ability to discriminate the existence of two WIMP species.

In the top left panel we learn that, compatibly with current direct detection limits in the parameter space of mass versus scattering cross section off nucleons, future detectors might be able to conclusively detect two WIMPs making up half of the cosmological dark matter each, as long as the mass of the lighter WIMP is below 40-50 GeV, and the heavier WIMP is lighter than 100 GeV or so. Also, the scattering cross section must not be much beyond current limits. The yellow shaded area indicates that while a good control on the background will not extend the discovery reach for a multi-WIMP scenario in scattering rate, it will indeed enlarge the detectable parameter space in mass, when current limits are considered. The green region in the figure points out that to detect two WIMPs might require scattering rates which are much larger (by roughly two orders of magnitude) than those needed to simply discover one of the dark matter particles.
The panel to the right highlights, for a fixed value of the cross section, the regions, in the $(m_1,m_2)$ parameter space, where discovery of a multi-particle dark matter sector is possible (the grey shaded area is excluded by current limits). There exist two general regions of parameter space where one can expect to discover and disentangle two WIMPs with a ton-sized Xenon detector: one corresponds to the lighter species having a mass around 10-20 GeV, and the second one in the 30-50 GeV range. The second region features a light mass for $m_2$ and extends in a wide range of masses for $m_1$ (which is assumed to have a relatively large scattering cross section, $10^{-44}\ {\rm cm}^2$). In short, a general requirement is that one of the two WIMPs be light (tens of GeV), and the second one sufficiently strongly interacting with matter.

The central panels explore the capabilities of a linear collider with a center of mass energy of 0.5 TeV. The contours reflect pretty obviously the requirements of kinematics: The blue region corresponds roughly to where $m_C<m_1$, and thus $\widetilde C$ cannot decay into either WIMP. In the green range, we approximately have $m_C<2\times m_1=m_2$, so that only $m_1$ is produced. When $m_C>m_1,\ m_2$ essentially the entire parameter space can be explored with a linear collider and the two stable WIMPs unambiguously discovered. In the right panel we point out that the only caveat to this situation is when either the two masses are very close to each other (i.e. small $\Delta$) or they are close to either $m_C$ or the kinematic reach of the collider (respectively, the upper left portion of the red contour, and the right edge). Still, a linear collider  would be potentially able to discover two stable WIMPs even if the relative mass splitting is of the order of a few percent (left edge of the red region).

For indirect detection, we indicate with dashed lines in the two lower panels the limits on monochromatic gamma-ray lines from current, pre-Fermi data, namely EGRET (at low masses) and atmospheric Cherenkov Telescopes (larger masses) \cite{Mack:2008wu, Pullen:2006sy}. We will give details of our assumptions in the relative section. The left panel shows that in this channel the prospects for singling out two gamma-ray lines originating from the monochromatic $\chi_{1,2}\chi_{1,2}\to\gamma\gamma$ channels are rather promising. The panel to the right highlights the existence of the ``sweet spot'' for Fermi-LAT in the parameter space of the pair annihilation cross sections, setting the first mass to 40 and the second to 80 GeV. For cross sections of the order of $10^{-25}\ {\rm cm}^3/{\rm s}$, and this combination of WIMP masses, Fermi-LAT can disentangle the existence of two monochromatic lines, with cross sections compatible with current gamma-ray data. The two green regions correspond to where we expect to detect either the lighter WIMP $\chi_1$ (upper green region) or the heavier WIMP $\chi_2$ (green region to the right).

\section{Direct Detection}\label{sec:direct}

A large class of current direct dark matter detection experiments measures the number $N$ of elastic collisions between WIMPs and target nuclei in a detector, per unit detector mass and per unit time, as a function of the nuclear recoil energy $E_r$. Given a local dark matter density $\rho_0 \simeq 0.3$ GeV cm$^{-3}$ (here with some fraction due to the first WIMP and another fraction due to the second one) and the velocity distribution $f(v_\chi)$ of WIMPs near the earth, the differential rate per unit detector mass and per unit time can be written as \cite{Bernal:2008zk}
\be \label{eq:diffrate}
\frac{dN}{dE_r} = \frac{\sigma_{\chi N} \rho_0}{2m_r^2m_\chi} F(E_r)^2 \int_{v_{min}(E_r)}^\infty \frac{f(v_\chi)}{v_\chi} dv_\chi ,
\ee
where the WIMP-nucleus cross section, $\sigma_{\chi N}$, is related to the WIMP-nucleon cross section, $\sigma_{\chi n}$, by $\sigma_{\chi N} = \sigma_{\chi n}(Am_r/M_r)^2$, with $M_r = \frac{m_\chi m_n}{m_\chi + m_n}$ the WIMP-nucleon reduced mass, $m_r = \frac{m_\chi m_N}{m_\chi + m_N}$ the WIMP-nucleus reduced mass, $m_\chi$ the WIMP mass, $m_n$ the nucleon mass, $m_N$ the nucleus mass, and $A$ the atomic weight. $F(E_r)$ is the Woods-Saxon form factor \cite{Engel:1991wq}
\be
F(E_r) = \frac{3j_1(qR_1)}{qR_1} e^{-\frac{1}{2}(qs)^2},
\ee 
where $q = \sqrt{2m_N E_r}$ is the momentum transferred, $j_1$ is a spherical Bessel function, $R_1 = \sqrt{R^2 - 5s^2}$, with $R \simeq 1.2 A^{1/3}$ fm, $A$ the mass number, and $s\simeq 1$ fm.
For the velocity distibution we take
\be
f(v_\chi)d^3v_\chi = \frac{1}{(v^0_\chi)^3 \pi^{3/2}} e^{-(v_\chi / v^0_\chi)^2} d^3v_\chi,
\ee
where $v^0_\chi \simeq 220$ km/s is the velocity of the Sun around the galactic center, and we have neglected the motion of the Earth around the Sun. After integrating over the angular part in order to find the speed distribution one gets
\be
f(v_\chi) dv_\chi = \frac{4 v^2_\chi}{(v^0_\chi)^3 \sqrt{\pi}} e^{-(v_\chi / v^0_\chi)^2} dv_\chi .
\ee
The integration over velocities in Eq.(\ref{eq:diffrate}) is limited, for a given recoil energy, to those that are kinematically allowed, meaning that there is a minimal velocity given by $v_{min}(E_r) = \sqrt{\frac{m_N E_r}{2 m_r^2}}$. From Ref.~\cite{Bernal:2008zk}, we take the background to be at the fixed value of $N^{bkg} = 1.2 \times 10^{-4}$ events/kg/day.

To asses whether a WIMP setup is or not detectable and whether we can discriminate between a single and a dual-component dark matter scenario, we proceed in the following way: we compute the differential scattering event rate (\ref{eq:diffrate}) for a given experiment and data acquisition time, for the two-component DM setup with the parameters of Eq.(\ref{eq:dirpar}). We then use a $\chi^2$ criterion to tell one of the three situations apart: (1) no detection: the fit to the background is better than the given confidence level; (2) detection, but no discrimination: the fit to the background is worse than the given confidence level; (3) detection and discrimination: the single best fit DM model does not give a good enough fit.

Let us explain in more detail how the $\chi^2$ criterion is used here. $N^{sign}$ is the signal, $N^{bkg}$ is the background, and $N^{obs} = N^{sign} + N^{bkg}$ is the total signal measured by the detector. We divide the recoil energy range between 4 and 32 keV in $n=7$ equidistant energy bins. We then compute three different $\chi^2$:
\begin{enumerate}
\item \textit{Fit using only the background with one parameter.}  
\begin{equation*}
\chi^2(N_b) = \sum_{i=1}^n \frac{(N_b - N_i^{obs})^2}{N_i^{obs}}
\end{equation*}
If the fit is better than a 95\% confidence level we declare no detection, otherwise we declare detection.
\item \textit{Fit using a single DM model with two parameters.}
\begin{equation*}
\chi^2(\sigma , m) = \sum_{i=1}^n \frac{( f_i(\sigma, m) +N^{bkg} - N_i^{obs})^2}{N_i^{obs}}
\end{equation*}
Here, the parameter $\sigma$ is the WIMP-nucleus cross section, and $f_i(\sigma, m)$ is the differential rate in Eq.(\ref{eq:diffrate}) integrated over the bin width, with $m$ the mass of the WIMP. Note that the background $N^{bkg}$ is held fixed. If this fit is better than a 95\% confidence level, we conclude that we cannot distinguish between the two species of dark matter, otherwise we declare discrimination.
\item \textit{Fit using a single DM model with three parameters.}
\begin{equation*}
\chi^2(\sigma , m, N_b) = \sum_{i=1}^n \frac{(f_i(\sigma, m) +N_b - N_i^{obs})^2}{N_i^{obs}}
\end{equation*}
The difference between this $\chi^2$ and the previous one is that the background has now become a parameter that we are going to fit. Again, we declare discrimination if this fit is worse than a 95\% confidence level.
\end{enumerate}

\begin{figure}[!th]
  \centering
 \includegraphics[width=125mm]{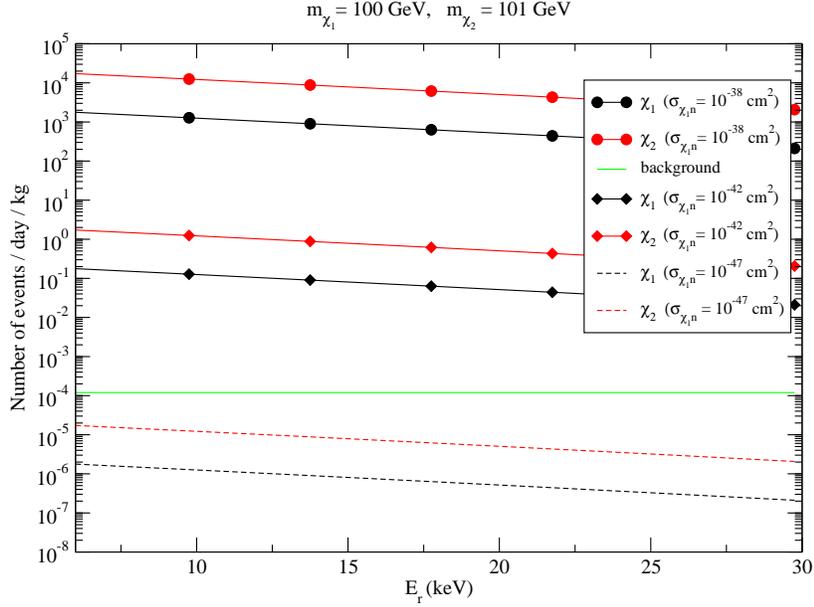}
  \caption{The number of events per kg per day is plotted as a function of the nuclear recoil energy on a logarithmic scale. For a small cross section (dashed lines), both signals for $\chi_1$ and $\chi_2$ are below the background and there is no detection. For a cross section of $10^{-42}$ cm$^2$ (diamonds) we can detect a signal, but the two lines, red and black, are too close to each other and we cannot distinguish between the two particles. For a large cross section (circles) we are way above the background, and now the two lines are far enough from each other to allow for discrimination.}
  \label{fig:glitch}
\end{figure}

To help visualize how we distiguish among (1) no detection, (2) detection, (3) discrimination, we plot in Fig. \ref{fig:glitch} the number of events measured, for different WIMP-nucleus cross sections, when the masses of $\chi_1$ and $\chi_2$ are very close.

In our study we consider a XENON-type experiment with 1000 kg of xenon and 5 years of data acquisition. In the following plots, the colors yellow and red both denote discrimination. In the yellow regions we use a chi-square test with two parameters, which corresponds to say that one has perfect control over the background; in the red regions we use a chi-square test with three parameters, one of which is the background itself.


\begin{figure}[hp]
  \centering
  \subfigure{
    \includegraphics[width=145mm]{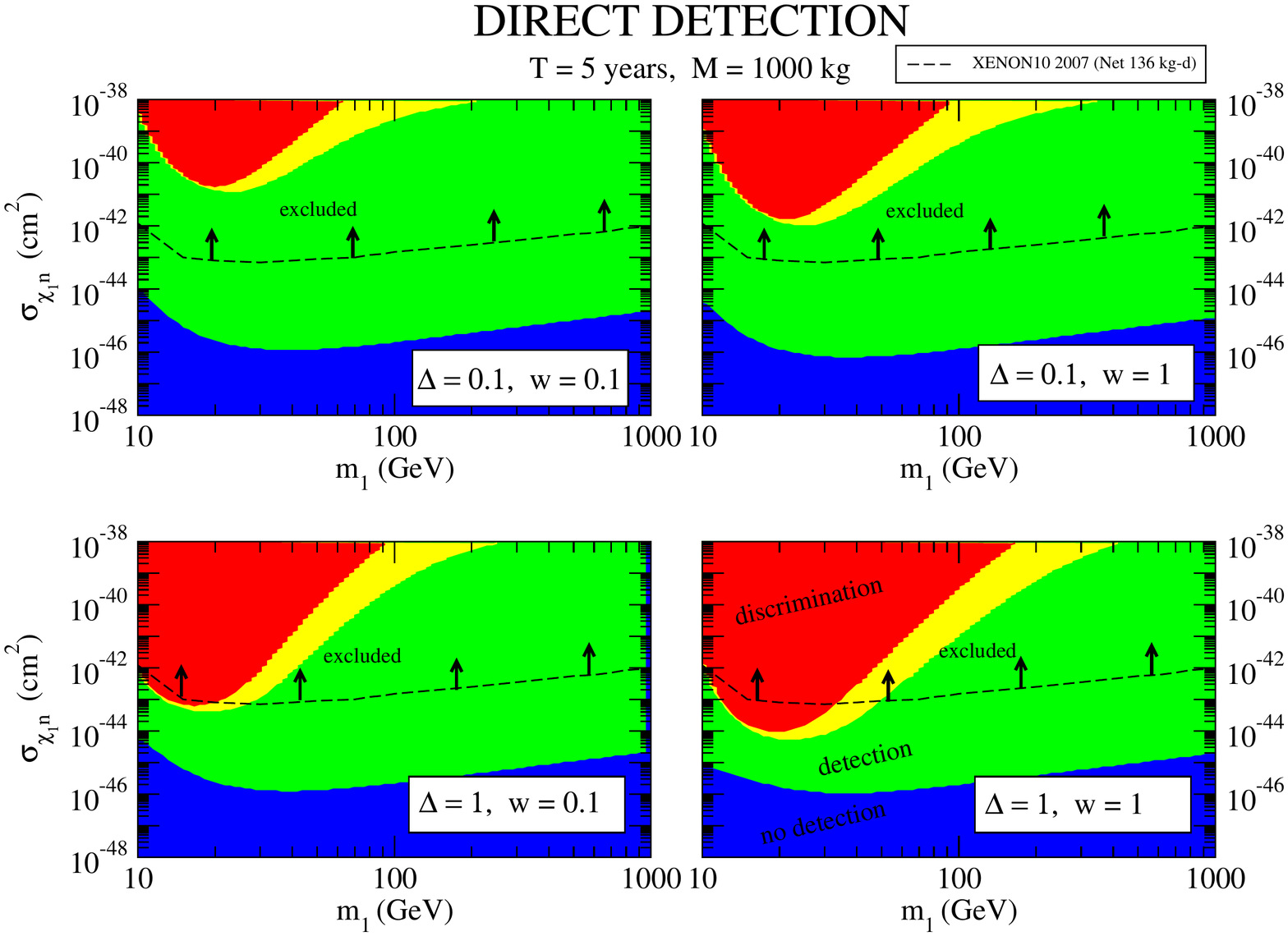}}\\[-.8cm]
  \subfigure{
    \includegraphics[width=145mm]{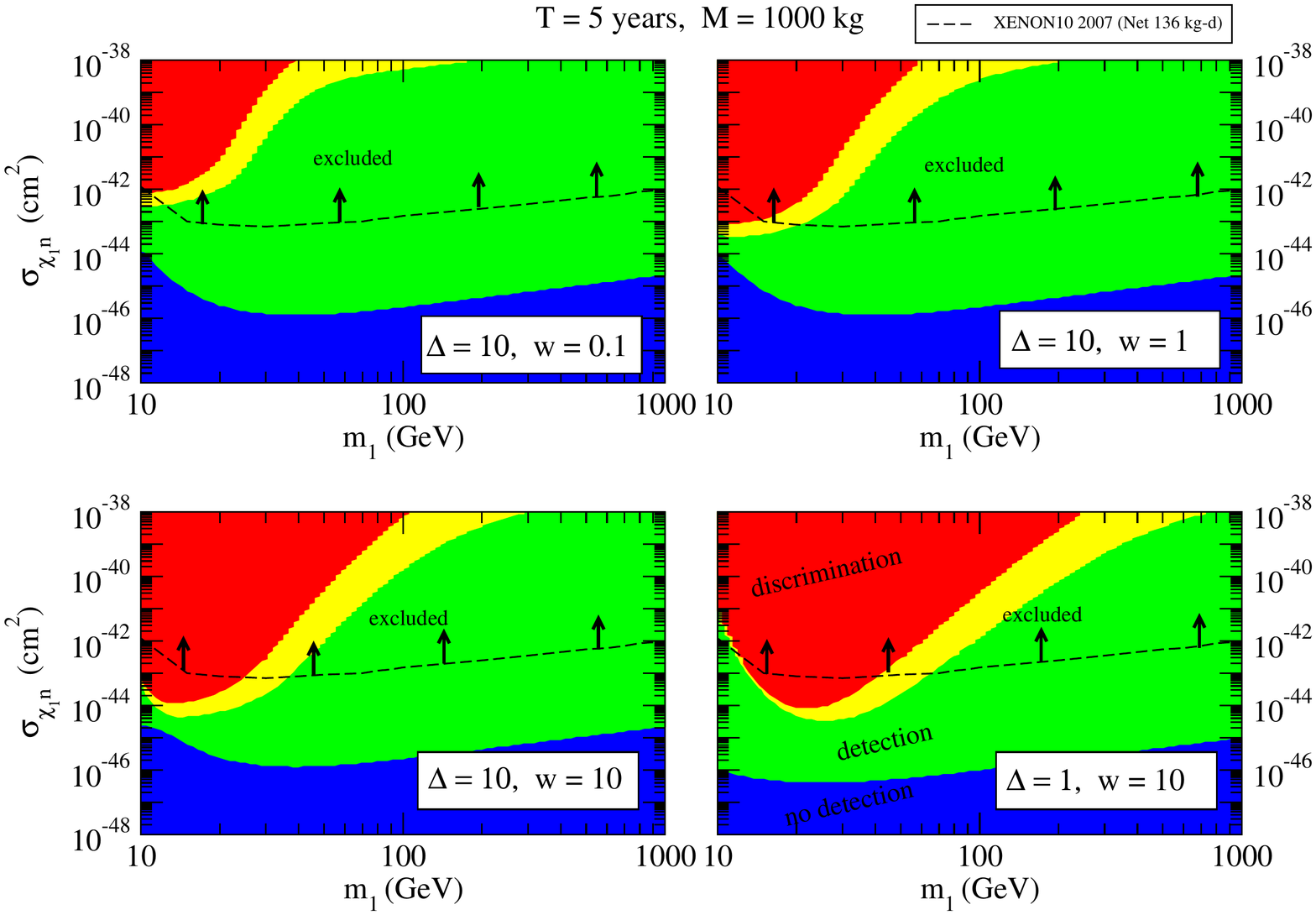}}\\[-1cm]
  \caption{\textit{Direct detection.} $\sigma_{\chi_1 n}$ vs $m_1$ is plotted at fixed values of $\Delta$ and $w$. The current limits from XENON10 2007 \cite{dendera} are also shown, with the arrows indicating the excluded region. {\em Color coding}: blue stands for {\em no detection}, green for {\em detection without discrimination}, red and yellow for {\em detection with discrimination}.}
  \label{fig:d5S1m1}
\end{figure}
Fig.~\ref{fig:d5S1m1} sets the stage showing the parameter space of the lighter WIMP mass, $m_1$, versus the WIMP-nucleon scattering cross section $\sigma_{\chi_1 n}$, for selected values of the second WIMP mass (we remind the reader that $m_2=m_1\times(1+\Delta)$) and of the scattering cross section off of nucleons for the second WIMP, set by the relation $\sigma_{\chi_2 n} = w\times\sigma_{\chi_1 n}(m_1/m_2)^2$, as in Eq.~(\ref{eq:wsigmasdir}). The various pair of values for $\Delta$ and $w$ we employ is specified in each panel, and we indicate the current limits with a dashed black line (the region excluded lies above the line).

We see that for direct detection, a necessary condition for discrimination (whether or not the background is well known) is to have a significant mass splitting between the WIMPs. For $\Delta=0.1$ (two top panels), discrimination is 2-3 orders of mangitude above the current exclusion limits. For $\Delta=1$, discrimination is possible with a good knowledge of the background and for light WIMPs, both in the 10-40 GeV range, with large scattering cross sections, around the current limits. Larger mass splittings ($\Delta=10$, third line of panels) entail detectability with discrimination only if the lighter WIMP is around $m_1\sim10$ GeV {\em and} the two dark matter candidates contribute comparably to the dark matter content of the Universe. Interestingly, since for direct dark matter detection the relevant particle density enters the detection rate linearly and not quadratically (as for the annihilation rate relevant for indirect detection), large values of $w$ (i.e. a setup where dark matter is dominated by the light component) can lead to discrimination of a multi-partite dark matter scenario. For a xenon detector, the best mass range for this case lies in the tens of GeV range for the lighter dark matter particle (larger values being favored with smaller mass splittings between the two WIMPs, see the bottom panels).



\begin{figure}[hp]
  \centering
  \includegraphics[width=172mm]{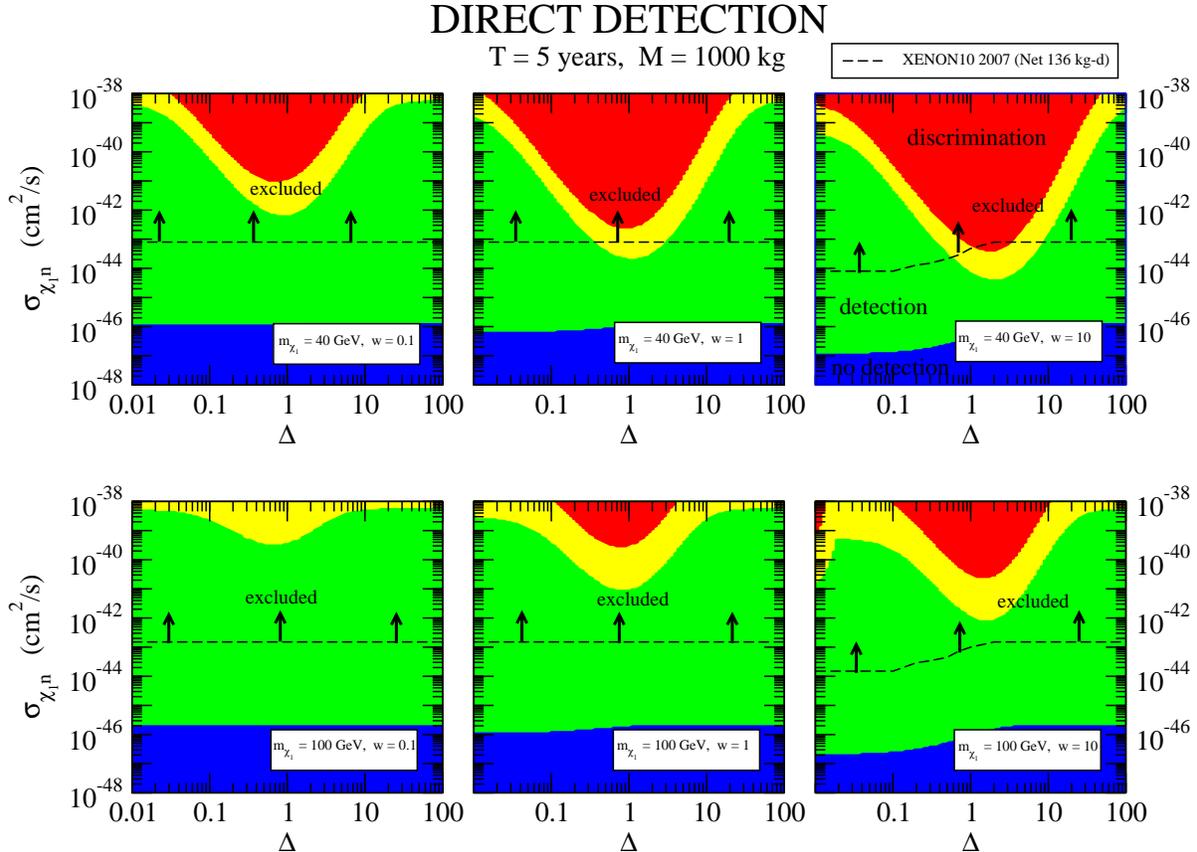}\\[-1.cm]
  \caption{\textit{Direct detection.} $\sigma_{\chi_1 n}$ vs $\Delta$ is plotted at fixed values of $m_1$ and $w$. The current limits from XENON10 2007 \cite{dendera} are also shown, with the arrows indicating the excluded region. {\em Color coding}: blue stands for {\em no detection}, green for {\em detection without discrimination}, red and yellow for {\em detection with discrimination}.}
  \label{fig:d5S1Delta}
\end{figure}

In Fig.~\ref{fig:d5S1Delta} we explore an alternative slice of parameter space, and investigate the prospects for discrimination on the $(\Delta,\sigma_{\chi_1 n})$ plane, for fixed values of $m_1=40$ GeV (upper panels) and 100 GeV (lower panels). The limits from current searches are also shown, and lie above the region where discrimination is possible only marginally, in the $m_1=40$ GeV case, for large $w=10$, when $\Delta$ is close to 1. In general, we find that the favored range for $\Delta$ for discrimination is always very peaked around 0.5-2, meaning that both a much heavier and a quasi-degenerate double-WIMP setup will be extremely hard to discover at direct detection: the best case scenario is for a mass ratio for the two WIMPs close to 2:1.

\begin{figure}[t]
  \centering
  \includegraphics[width=172mm]{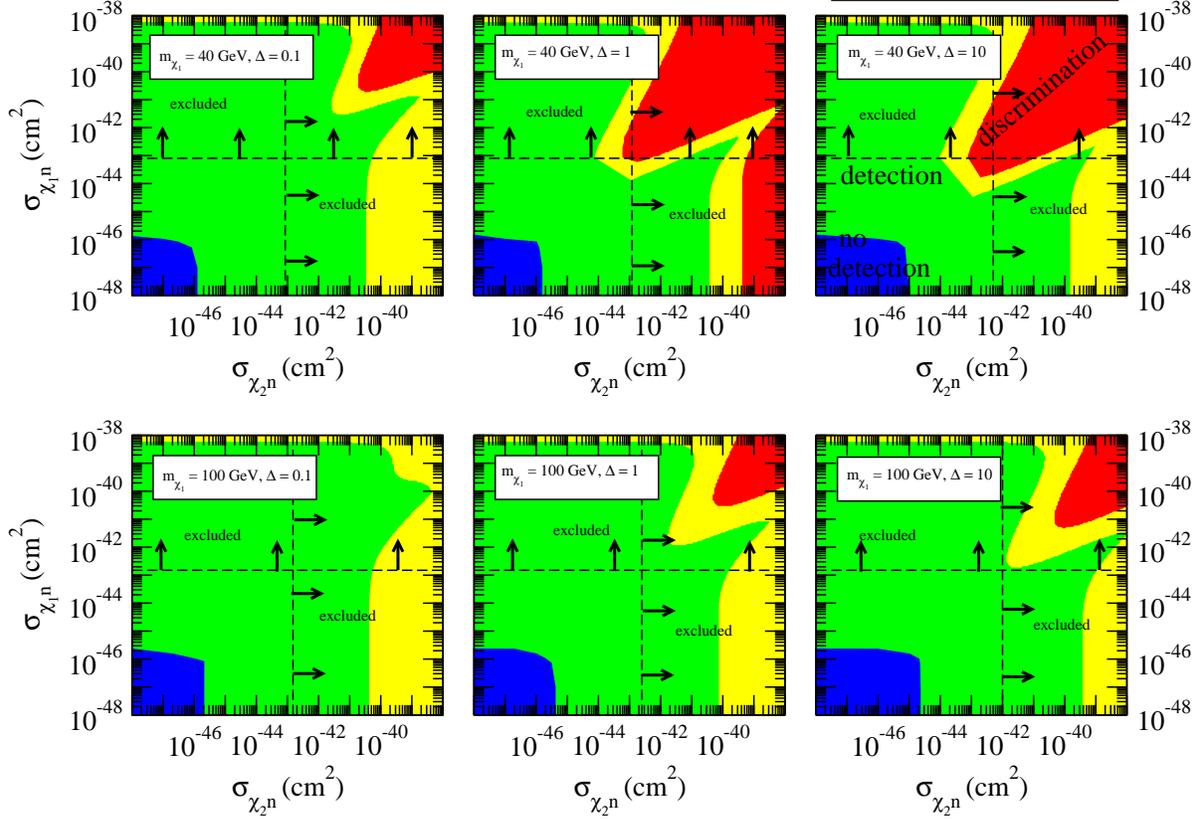}\\[-1.cm]
  \caption{\textit{Direct detection.} Here we relax the assumption (\ref{eq:wsigmasdir}) and we plot $\sigma_{\chi_1 n}$ vs $\sigma_{\chi_2 n}$ at fixed values of $m_1$ and $\Delta$. The current limits from XENON10 2007 \cite{dendera} are also shown, with the arrows indicating the excluded region. {\em Color coding}: blue stands for {\em no detection}, green for {\em detection without discrimination}, red and yellow for {\em detection with discrimination}.} 
  \label{fig:d5S1S2}
\end{figure}

In Fig.~\ref{fig:d5S1S2} we relax the assumption of Eq.~(\ref{eq:wsigmasdir}) and study the ranges of WIMP-nucleon scattering cross sections that can be detected, for fixed values of the masses. In the upper panels the lighter WIMP weighs 40 GeV and the second one has a relative mass splitting of $\Delta=0.1$, 1 and 10, left to right. The second row has the same splittings, but for a light WIMP of 100 GeV. The panels highlight that the most favorable situation is either if one of the two WIMPs has a very large scattering rate compared to the second (the vertical bands to the right and at the top of all panels), or if the two scattering cross sections are very close to each other and large enough (the wedge in the upper right corners). The size of the yellow regions highlights the importance of background suppression and control for the discovery of a multi-partite dark matter setup.

\begin{figure}[t]
  \centering
   \includegraphics[width=145mm]{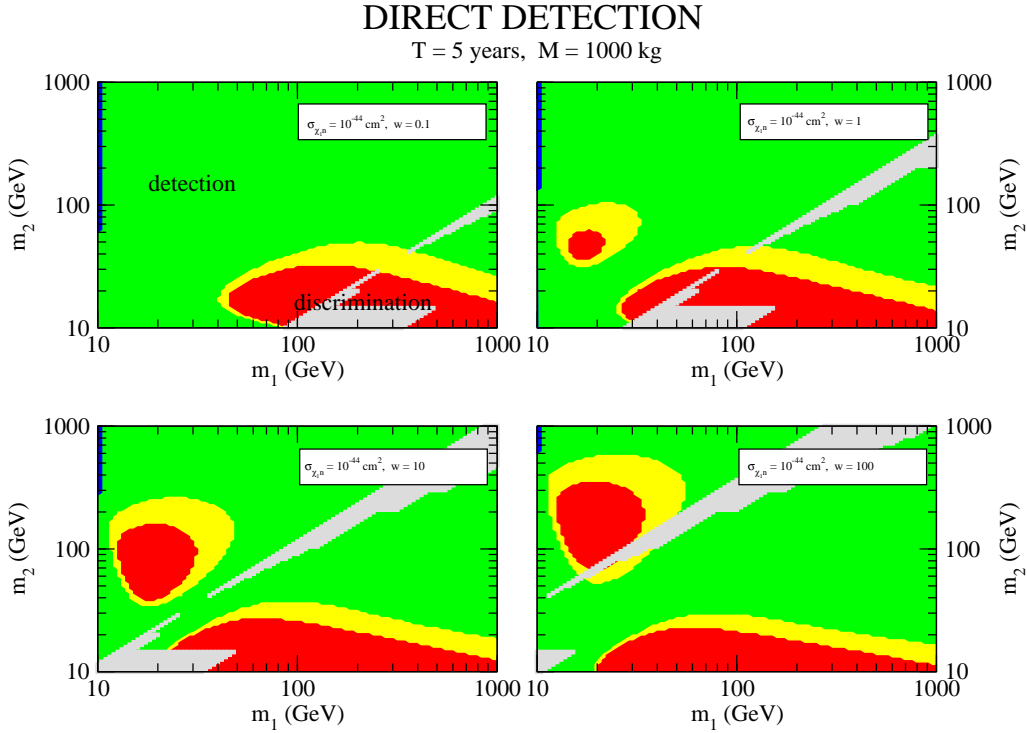}\\[-1.cm]
  \caption{\textit{Direct detection.} Here we replace $\Delta$ with $m_2$ in the parameter space (\ref{eq:dirpar}) and plot $m_2$ vs $m_1$ at fixed values of $\sigma_{\chi_1 n}$ and $w$. The shaded area is the region excluded by the limits from XENON10 2007 \cite{dendera}. {\em Color coding}: blue stands for {\em no detection}, green for {\em detection without discrimination}, red and yellow for {\em detection with discrimination}.} 
  \label{fig:d5m1m2}
\end{figure}

As evident from Figs.~\ref{fig:d5S1Delta} and \ref{fig:d5S1S2}, the viability of a multi-component dark matter scenario discovery at direct detection experiments depends very sensitively on the choice of masses, and although it also depends on the relative mass splitting and relative contribution to the dark matter density, the precise value of the WIMP masses appears to be a crucial factor. We explore this fact in Fig.~\ref{fig:d5m1m2}, where we plot $m_2$ vs $m_1$ (we shade in grey the regions ruled out by current experiments \cite{dendera}). We take $\sigma_{\chi_1 n}=10^{-44}\ {\rm cm}^2$, and vary $w=0.1$, 1, 10 and 100.

We find that discrimination is possible if one of the two WIMPs is light, namely with a mass below 50 GeV, and the second WIMP has a mass between 20 GeV and 1 TeV. These ranges depend on the relative abundance $w$, but the maximal extension of the regions outside the current experimental bounds is found for large $w$ (i.e. for a Universe dominated, in terms of relative matter abundance, by the lighter WIMP). Notice that the discovery regions in yellow and red are not symmetric in the plot because of the assumed scaling of $\sigma_{\chi_2 n}$ with $w$ and the WIMP masses.

\begin{figure}[t]
  \centering
  \includegraphics[width=172mm]{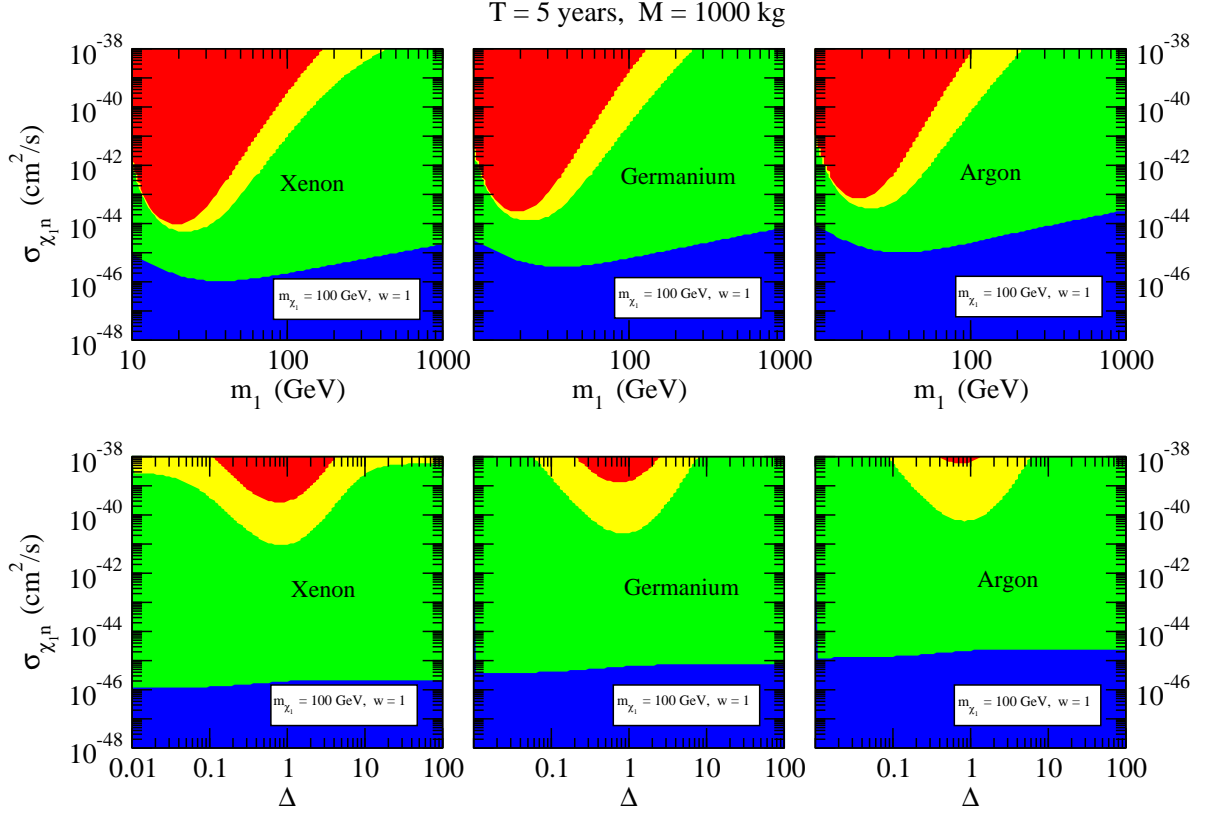}\\[-1.cm]
  \caption{\textit{Direct detection.} We compare three experiments running for the same exposure time and using the same mass of xenon, germanium and argon respectively. {\em Color coding}: blue stands for {\em no detection}, green for {\em detection without discrimination}, red and yellow for {\em detection with discrimination}.} 
  \label{fig:nuclei}
\end{figure}

Fig.~\ref{fig:nuclei} explores the different prospects for different {\em target nuclei}. Specifically, we compare three experiments with the same exposure time and target mass, but with xenon, germanium and argon nuclei (from left to right), in the planes defined by the lighter WIMP mass and scattering cross section (upper panels) and $\Delta$ versus again the scattering cross section (bottom panels). Although with naive assumptions on the background rate the heavier the nucleus the better the discrimination prospects, we do not find any evidence that the discovery of a multi-partite dark matter is favored with one target nucleus choice versus another. The preferred mass and relative mass splitting for discrimination  is also target-independent.

\section{Collider Searches}\label{sec:colliders}

The new energy frontier that the Large Hadron Collider will soon explore will surlely have a tremendous impact on our understanding of physics beyond the Standard Model. As history teaches us, hadronic machines often correspond to ``discovery'' apparata, as opposed to leptonic colliders, where the goal has often been to attain precision measurements. In an optimistic scenario, physics results and discoveries with the LHC will trigger the international community to move on with a new generation of precision colliders, which many have envisioned as a natural next step, in the form of an international linear electron-positron collider (ILC).

As discussed above, in the present model-independent setup, we specify neither the specific particle physics model for the stable WIMPs nor the broader particle content of the model. The latter is of crucial importance to specify the decay channels and resulting observational signatures of new heavy particles produced at the LHC. Although studies have addressed this problem in the context of canonical collider signatures, including large transverse missing energy, jets and high energy leptons, where mass reconstruction of the produced particles can also be achieved (for a recent study see e.g.~\cite{Matchev:2009iw}), it is very problematic to set this problem in model-independent terms. We therefore focus on a simple scenario that an ILC might face: one where a massive charged particle species $\widetilde C^\pm$ is pair-produced in $e^+e^-$ collisions, and it decays into the two stable WIMPs with a certain branching ratio, giving off a high transverse-momentum lepton (for simplicity we assume the latter to be an electron or a positron). We showed schematic relevant Feynman diagrams in Figure~\ref{fig:diagrams}.
As for indirect detection, we assume there is no cross-channel where for instance the charged particle decays into both stable WIMPs, plus other standard model particles. This channel would anyway be kinematically suppressed.

In this context, the key parameter of the game is the mass of the charged species $\widetilde C^\pm$, which will then determine the production cross section at the ILC\be
\sigma_{\widetilde C^+\widetilde C^-}\propto \frac{\alpha^2}{s}\sqrt{1-\frac{4m_{\widetilde C}^2}{s}}\, .
\ee
 Here we take the center of mass energy to be $\sqrt{s} = 500$ GeV, as might be expected for the early phase of an ILC, and set $\alpha = \alpha (m_Z) = 1/129$. In the context of e.g. supersymmetry, the constant of proportionality in the above expression would be $\frac{\pi}{2} q^2_{\widetilde C}$, where $q_{\widetilde C}$ is the charge of $\widetilde C$. To be conservative, we take here such a constant to be 1 in our analysis.  
The additional parameters are the decay widths into the two stable species, plus light leptons (the case of quarks would be similar, in case $\widetilde C$ is a strongly interacting species):
\be
\Gamma_{1,2}=\alpha_{1,2}m_{\widetilde C}\left(1-\frac{m^2_{\chi_{1,2}}}{m_{\widetilde C}^2}\right)\, .
\ee
As in the case of direct detection (and as discussed in Sec.~\ref{subsec:collider}), we can relate the $\Gamma_{1,2}$ to $r_{\alpha^2} \simeq w$, assuming that the particle that dominantly mediates the pair annihilation cross section into Standard Model particles is ${\widetilde C}$ (for instance, in the case of supersymmetry, a squark mediating the pair annihilation of neutralinos into a quark-antiquark pair in a $t$-channel). With the further simplifying assumption that $\widetilde C$ only decays into $\chi_1$ or $\chi_2$, we have the branching ratios
\be
BR_{1,2} = \frac{\Gamma_{1,2}}{\Gamma_1 +\Gamma_2}\, .
\ee
Assuming that the cross sections are inversely proportional to the the dark matter energy densities we have $\alpha^2_{1,2} \propto \sigma_{1,2} \propto 1/\Omega_{1,2}$, and, using (\ref{eq:w}), we can write
\bea
BR_1 &=& \frac{1}{1+\sqrt{w}\frac{m^2_{\widetilde C} - m^2_2}{m^2_{\widetilde C} - m^2_1}} \label{eq:br1} \\
BR_2 &=& \frac{1}{1+\frac{1}{\sqrt{w}}\frac{m^2_{\widetilde C} - m^2_1}{m^2_{\widetilde C} - m^2_2}}.  \label{eq:br2}
\eea
The number of signal events is
\be
N_{1,2} = BR_{1,2} N\, ,
\ee
where
\be
N=f_{\rm eff}\times \sigma_{\widetilde C^+\widetilde C^-} \times {\cal L}
\ee
and we have the following estimates for the ILC parameters \cite{privcom}
\bea
f_{\rm eff} &=& 0.1 \qquad \rm{efficiency \ factor}\, , \\
{\cal L} &=& 100 \ \rm{fb}^{-1} \qquad \rm{integrated \ luminosity}.
\eea
We declare detection when either $N_1 > 20$ or $N_2>20$.

As for the discrimination of two distinct stable WIMP species in the final state, we consider the accuracy of ILC mass measurements, given the number of events and the intrinsic detector resolution:
\be
\frac{\Delta m_{1,2}}{m_{1,2}}\sim {\rm Max}\left(\left(\frac{\Delta m_{1,2}}{m_{1,2}}\right)_{\rm max},\frac{a}{\sqrt{N_{1,2}}}\right)
\ee
From \cite{Yang, ilcdesign, lcworkshop} we get the following estimates:
\bea 
\left(\frac{\Delta m_{1,2}}{m_{1,2}}\right)_{\rm max} &=& 0.0005 \\
a &=& 0.11
\eea
where $a$
indicates the statistically-limited detector mass resolution, as
estimated in the simulations presented in Ref.~\cite{Yang}. Specifically, the
simulations in \cite{Yang} present the relative error in the determination of
the smuon mass for SPS point \#1 \cite{sps} at a 500 GeV center of mass
linear
collider, including the effects of initial state radiation,
beamstrahlung and beam energy spread.
We then declare discrimination when both $\chi_1$ and $\chi_2$ are detectable and
\be
m_1 + 5\Delta m_1 < m_2 - 5\Delta m_2
\ee
\textit{i.e.} the two masses are discernible at $5\sigma$.


\begin{figure}
  \centering
  \subfigure{
    \includegraphics[width=145mm]{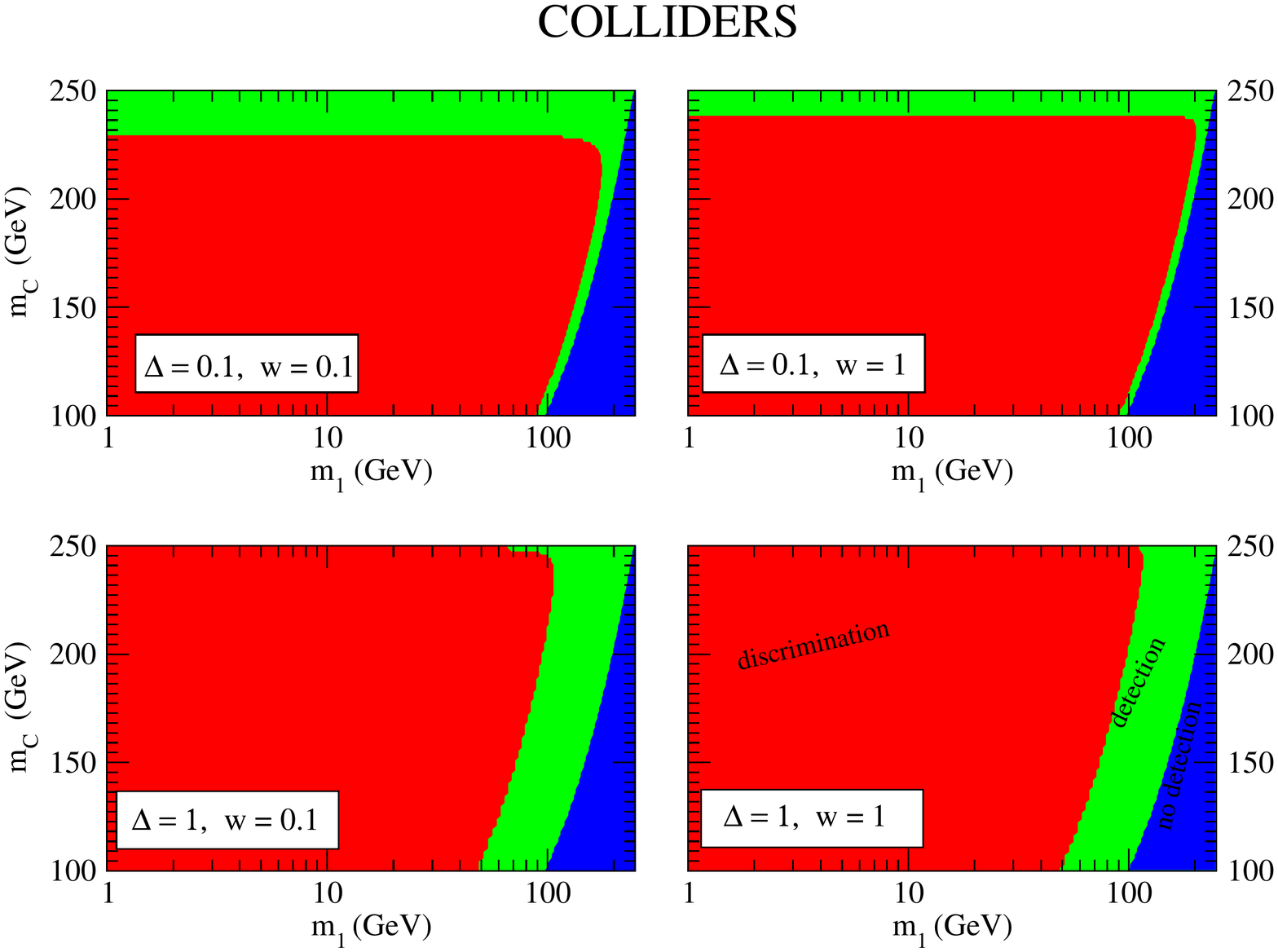}}\\[-.8cm]
  \subfigure{
    \includegraphics[width=145mm]{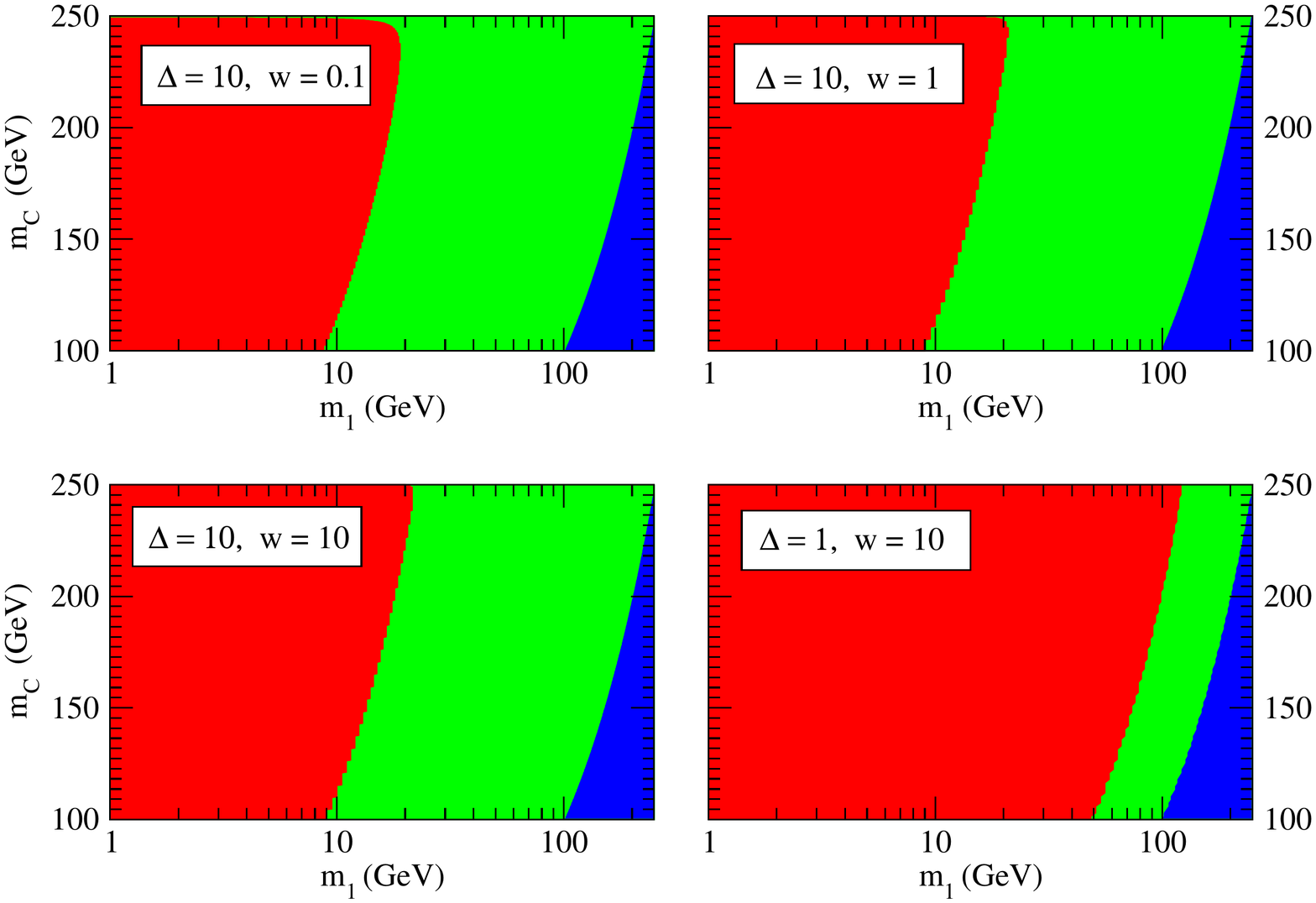}}\\[-1.cm]
  \caption{\textit{Collider searches.} $m_{\widetilde C}$ vs $m_1$ is plotted at fixed values of $\Delta$ and $w$. {\em Color coding}: blue stands for {\em no detection}, green for {\em detection without discrimination}, red for {\em detection with discrimination}.}
  \label{fig:cmcm1}
\end{figure}

Fig.~\ref{fig:cmcm1} illustrates the reach of an ILC with the present setup, in the $(m_1,m_C)$ parameter space, for fixed values of the mass splitting $\Delta$ and relative dark matter species abundance $w$. While the blue edges essentially trace the kinematically defined parameter space region where $m_C>m_1$, the red contours correspond to satisfying both the requirement that $m_C>m_2$ and that a sufficiently large event statistics is achieved. In particular in our scenaro, for small $w$ (for a Universe dominated by the heavier species) the heavier WIMP interacts more weakly (since it annihilates less efficiently) and $\widetilde C$ has a consequently smaller probability to produce it in the decay mode. This is also evident from Eq.~(\ref{eq:br2}) above. Comparing the upper two panels this is very easily seen: if $w=0.1$ the charged particle must be light enough as to have a large enough number of events leading to the heavier WIMP in the final state (left panel). When $w=1$, instead, the reach extends almost all the way to the kinematic limit. The other panels illustrate the same point: A smaller value of $w$ (or a very large one too, see the bottom right panel with $w=10$) indicates that one of the two stable species is the preferred decay mode for $\widetilde C$, and it is harder to have enough statistics to achieve discrimination. If both particles are kinematically open final states, the $w\sim1$ scenario is the most favorable one at an ILC.

\begin{figure}
  \centering  
  \includegraphics[width=172mm]{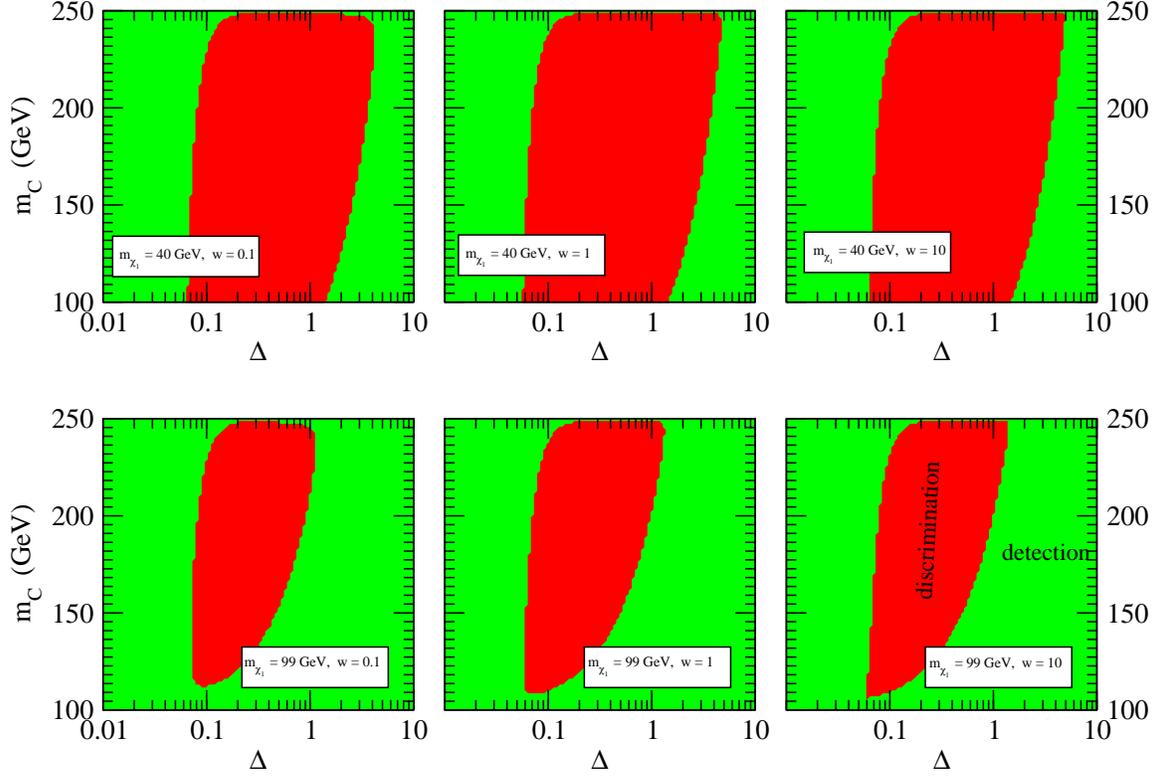}\\[-1.cm]
  \caption{\textit{Collider searches.} $m_{\widetilde C}$ vs $\Delta$ is plotted at fixed values of $m_1$ and $w$. {\em Color coding}: green stands for {\em detection without discrimination}, red for {\em detection with discrimination}.} 
  \label{fig:cmcDelta}
\end{figure}

Fig.~\ref{fig:cmcDelta} studies, for fixed values of the lighter WIMP mass (40 GeV and 99 GeV in the upper and lower panels, respectively) and for $w=0.1$, 1 and 10, the regions of the $(\Delta,m_C)$ parameter space where discrimination of two stable WIMPs is possible. Discovery is possible (green region) throughout the entire parameter space shown. The right (and lower, for $m_1=99$ GeV) edges of the red regions where the discrimination of two stable WIMP species is possible are dictated by requiring that the heavier WIMP be lighter than the charged particle, while those to the left depend upon the finite detector mass resolution: the best possible case ($w\sim1$) suggests that an ILC might be able to discover two distinct stable WIMPs as long as their relative mass splitting is larger than 5-10\%. The upper edges again are determined by the requirement of having a large enough statistics. The take-away message from Fig.~\ref{fig:cmcDelta} is that the ideal relative mass splitting to single out two stable WIMPs at an ILC ranges between $0.1\lesssim\Delta\lesssim2$.

\begin{figure}
  \centering  
  \includegraphics[width=172mm]{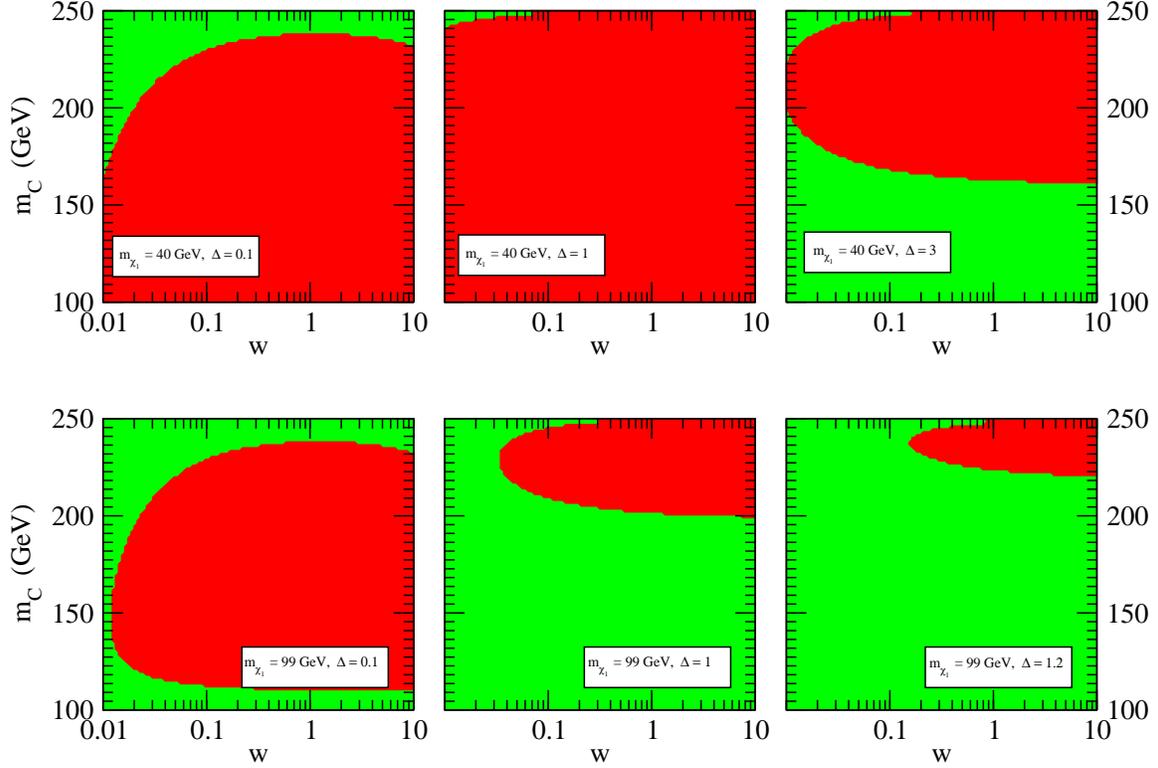}\\[-1.cm]
  \caption{\textit{Collider searches.} $m_{\widetilde C}$ vs $w$ is plotted at fixed values of $m_1$ and $\Delta$. {\em Color coding}: green stands for {\em detection without discrimination}, red for {\em detection with discrimination}.} 
  \label{fig:cmcw}
\end{figure}

We further study the role of $w$ in Fig.~\ref{fig:cmcw}, where we study discrimination of a dual-component stable WIMP scenario in the $(w,m_C)$ plane. The figure highlights that the ideal situation for discrimination corresponds to comparable decay branching ratios into the two stable species. The shape of the panels to the right, with larger $\Delta$, is determined by the requirement of producing both WIMPs (e.g. in the lower middle panel, the second WIMP has a mass of around 200 GeV, hence the red region starts where $\widetilde C$ can actually decay into it). Again, when $m_C\to250$ GeV, the kinematic reach of the collider is reached, and the total number of events decreases accordingly, leading to worse discrimination performance.

\begin{figure}
  \centering
    \includegraphics[width=172mm]{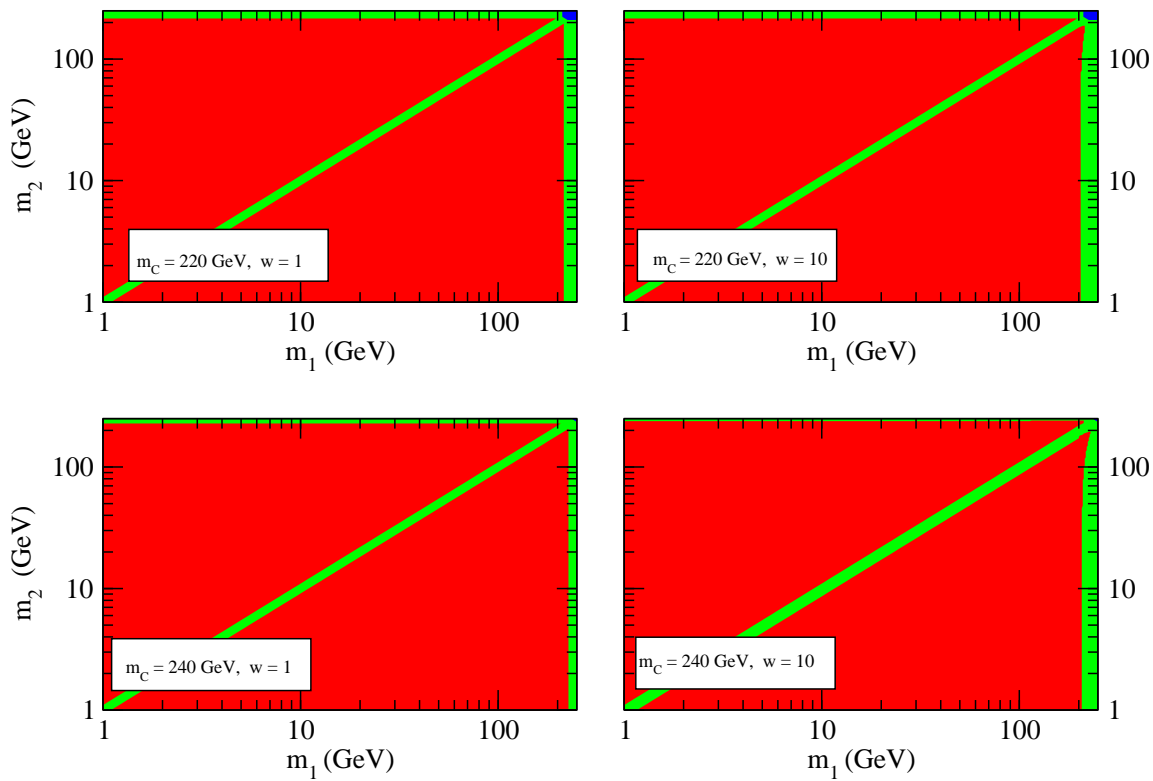}\\[-1.cm]
  \caption{\textit{Collider searches.} Here we replace $\Delta$ with $m_2$ in the parameter space (\ref{eq:collpar}) and plot $m_2$ vs $m_1$ at fixed values of $m_{\widetilde C}$ and $w$. The asymmetry in the 4th plot comes from the combination of $w$ and the masses in Eq. (\ref{eq:br1}) and (\ref{eq:br2}). {\em Color coding}: green stands for {\em detection without discrimination}, red for {\em detection with discrimination}.}
  \label{fig:cm2m1}
\end{figure}
In the last figure of this section, Fig.~\ref{fig:cm2m1}, we study the plane defined by the two stable WIMP masses $m_1$ and $m_2$, for fixed values of $w$ and of $m_C$. As expected, discrimination is problematic in the region of parameter space where $m_1\sim m_2$. In addition, the right edge of the plots to the right illustrates the critical issue of a largely asymmetric distribution for the relative dark matter densities. When dark matter is dominated by the first WIMP species ($w=10$), we see from Eq.~(\ref{eq:br1}) that $\chi_1$ is only very rarely produced in the decay chain of $\widetilde C$. The branching ratio $BR_1$ is even more suppressed if $m_1\to m_{\widetilde C}$. Therefore, we see in the plots that at constant values of $m_2$ there is some upper limit to $m_1$; beyond this limit we would not detect $\chi_1$ anymore because the branching ratio $BR_1$ would be too small.

\section{Indirect Detection}\label{sec:indirect}
The indirect detection of dark matter can proceed in a number of channels, or ``messengers'' \cite{multimessanger}, including but not limited to antimatter, gamma rays and high-energy neutrinos (for reviews in the context of several particle physics models for dark matter see e.g. \cite{Jungman:1995df,Bergstrom:2000pn,Bertone:2004pz,Hooper:2007qk}). The detection of gamma rays produced in the annihilation of dark matter, or in the radiation emitted in the subsequent energy losses of the energetic electrons and positrons (e.g. via Inverse Compton up-scattering of photons in the Cosmic Microwave Background) has been discussed at length in the literature (see e.g. the review of the possible dark matter search strategies with Fermi-LAT given in Ref.~\cite{fermilatdm}). 

In order to discover a multi-component dark matter scenario, we focus our attention on the detection of multiple monochromatic lines arising from the direct annihilation of dark matter pairs into two photons. This poses a fundamental problem. Additional lines can emerge from other channels, such as dark matter pair-annihilation into a photon plus a neutral gauge or Higgs Boson \cite{Ullio:1997ke}, or in the context of universal extra-dimensional models \cite{Hooper:2007qk, Bertone:2009cb}. Thus, given two lines for instance, one cannot distinguish whether the second line is due to a second WIMP or is just an additional line arising from the first WIMP (via annihilation into $Z\gamma$ or $H\gamma$ for example). Because of this caveat, indirect detection cannot be used to claim a discovery of multi-component dark matter, but it could provide a valuable verification if evidence for such a scenario is found via direct detection and/or collider searches. Therefore, it is useful to carry out the analysis here and compare to the results obtained from the other channels in the previous sections.

The signal flux of monochromatic photons for the $\chi_1\chi_1\to\gamma\gamma$ and for the $\chi_2\chi_2\to\gamma\gamma$ can be cast as \cite{Cesarini:2003nr}
\be\label{eq:signal1}
\phi_{\chi_{1}}=\frac{7.5\times10^{-10}}{{\rm cm}^2\ {\rm s}}\left(\frac{w}{1+w}\right)^2\left(\frac{\langle\sigma_{\chi_{1}\chi_{1}\to\gamma\gamma}v\rangle}{10^{-26}\ {\rm cm}^3\ {\rm s}^{-1}}\right)\left(\frac{50\ {\rm GeV}}{m_{1}}\right)^2\left(\frac{\Delta\Omega}{\rm sr}\right)J\left(\Delta\Omega\right)
\ee
and
\be \label{eq:signal2}
\phi_{\chi_{2}}=\frac{7.5\times10^{-10}}{{\rm cm}^2\ {\rm s}}\frac{w}{(1+w)^2}\left(\frac{\langle\sigma_{\chi_{1}\chi_{1}\to\gamma\gamma}v\rangle}{10^{-26}\ {\rm cm}^3\ {\rm s}^{-1}}\right)\left(\frac{50\ {\rm GeV}}{(1+\Delta)m_1}\right)^2\left(\frac{\Delta\Omega}{\rm sr}\right)J\left(\Delta\Omega\right),
\ee
where
\be\label{eq:deltaom}
J\left(\Delta\Omega\right)\equiv\frac{1}{\Delta\Omega}\frac{1}{8.5\ {\rm kpc}}\left(\frac{1}{0.3\ {\rm GeV}{\rm cm}^{-3}}\right)^2\int_{\Delta\Omega}{\rm d}\Omega\int_{\rm l.o.s}{\rm d}l(\Omega)\left(\rho^2(l)\right)
\ee
is the integrated line-of-sight dark matter density squared, over an angular region $\Delta\Omega$, in a given direction in the sky. Given that the monochromatic two-photon line is typically a statistically-limited indirect detection channel, the best channels include those involving observations with a large number of collected photons. Possible choices are the extragalactic diffuse signal from all redshifts and all halos, the galactic diffuse signal, the galactic center region and extragalactic targets such as local dwarf galaxies and nearby clusters of galaxies. We consider here as a case-study the Milky Way Galactic Center, which, for density profiles concentrated in the central regions of the Galaxy, can lead to the best signal-to-noise among the possible dark matter search targets listed above (see e.g. \cite{galactic_center}). 

We assume here a fiducial NFW profile \cite{Navarro:1996gj}, featuring a local dark matter density at the Sun position of $0.3\ {\rm GeV}{\rm cm}^{-3}$, a scaling radius of $r_s=20$ kpc, and the high-energy Fermi-LAT angular resolution $\Delta\Omega=3.83\times 10^{-3}$ sr, which corresponds to an angle of $\psi = 2$ degrees within the direction of the Galactic Center. We obtain the following:
\be
\left(\frac{\Delta\Omega}{\rm sr}\right)J\left(\Delta\Omega\right)\simeq2.30 \qquad (\Delta\Omega=3.83 \times 10^{-3}\ {\rm sr}).
\ee
As far as the background is concerned, we refer the reader to the detailed analysis of Ref.~\cite{galactic_center}; here we take a subset of the background sources mentioned in that study, and consider contributions from (1) the H.E.S.S. detected source at the Galactic Center, (2) the EGRET source near the GC, (3) diffuse radiation, divided into a galactic and an extragalactic component. Specifically, we use the following background fluxes \cite{Dodelson:2007gd}:
\bea
\phi^{\rm bckg}_{\rm HESS}&=&10^{-8}\left(\frac{E_\gamma}{\rm GeV}\right)^{-2.25}\ {\rm GeV}^{-1}\ {\rm cm}^{-2}\ {\rm s}^{-1},\\
\phi^{\rm bckg}_{\rm EGRET}&=&2.2\times 10^{-7}\left(\frac{E_\gamma}{\rm GeV}\right)^{-2.2}\times\exp\left(-E_\gamma/30\ {\rm GeV}\right)\ {\rm GeV}^{-1}\ {\rm cm}^{-2}\ {\rm s}^{-1}
\eea
and, from \cite{Serpico:2008ga}
\bea
\phi^{\rm bckg}_{\rm diff\_gal}&=&1186\times 10^{-6}\left(\frac{E_\gamma}{\rm GeV}\right)^{-3}\ {\rm GeV}^{-1}\ {\rm cm}^{-2}\ {\rm s}^{-1},\\
\phi^{\rm bckg}_{\rm diff\_ex}&=&3.66\times 10^{-6}\left(\frac{E_\gamma}{0.451\ {\rm GeV}}\right)^{-2.1}\left(\frac{\Delta\Omega}{\rm sr}\right)\ {\rm GeV}^{-1}\ {\rm cm}^{-2}\ {\rm s}^{-1}.
\eea

We consider an effective area for Fermi-LAT of $8500\ {\rm cm}^2$, which is appropriate for the range of energy we are dealing with (see {\em e.g.} \cite{LATperformance}), and an exposure time corresponding to 5 years with the GC within the Fermi-LAT field of view 50\% of the time \cite{GLASTsysterr}. The product $A_{\rm eff}\times t_{\rm obs}$ is therefore $\approx 6.7\times 10^{11}\ {\rm cm}^2\ {\rm s}$. Finally, we consider an energy resolution $\Delta E_\gamma/E_\gamma\simeq0.1$. 

Given a monochromatic flux of photons at a certain energy $E_{0}$, we compute the number of background events $N_{\rm bckg}$ in the bin between $0.95E_0$ and $1.05E_0$, and compare it with the number of signal events $N_{\rm sig}$, computed from Eqs.~(\ref{eq:signal1}) and (\ref{eq:signal2}). We then declare that a given dark matter setup is {\em detectable} if, for either $\chi_1$ or $\chi_2$, the following two conditions are satisfied:
\bea
\frac{N_{\rm sig}}{\sqrt{N_{\rm bckg}}}&>&5\\
N_{\rm sig}&>&50
\eea
The two conditions above require a signal-to-noise (the latter assumed to be gaussian) at least at the $5-\sigma$ level, and a total number of signal photon events larger than 50. We instead declare to have {\em discrimination} for a setup where (1) both $\chi_1$ and $\chi_2$ are detectable and (2) the mass determination is such that 
\be 
m_1+5\Delta m_1<m_2-5\Delta m_2,
\ee
{\em i.e.} the two masses are descernible at 5$\sigma$. We estimate 
\be
\Delta m_i=\frac{0.1}{\sqrt{N_{\rm sig}}}m_i
\ee


\begin{figure}[!ht]
  \centering
  \subfigure{
    \includegraphics[width=150mm]{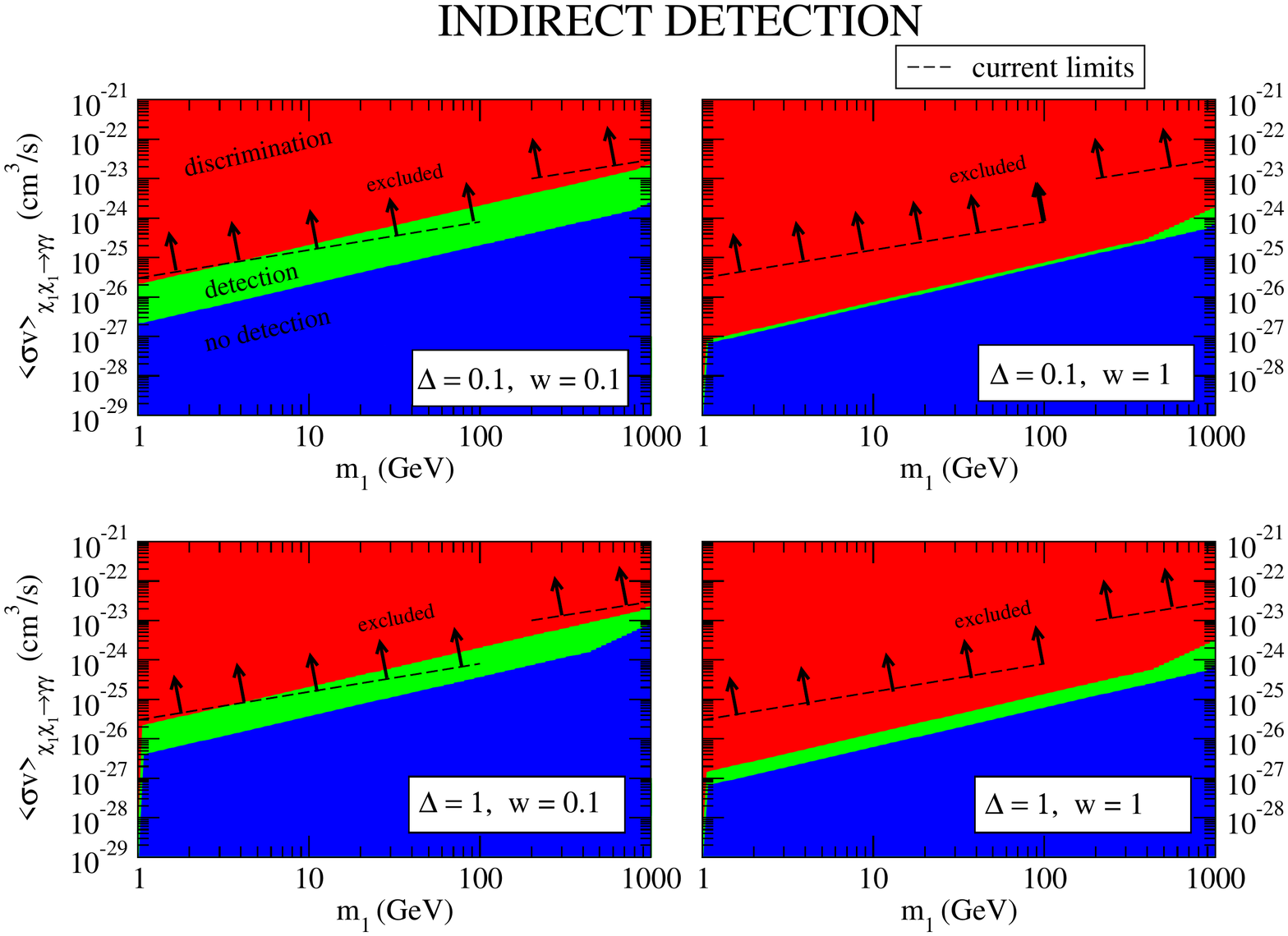}}\\[-1cm]
  
  \subfigure{
    \includegraphics[width=150mm]{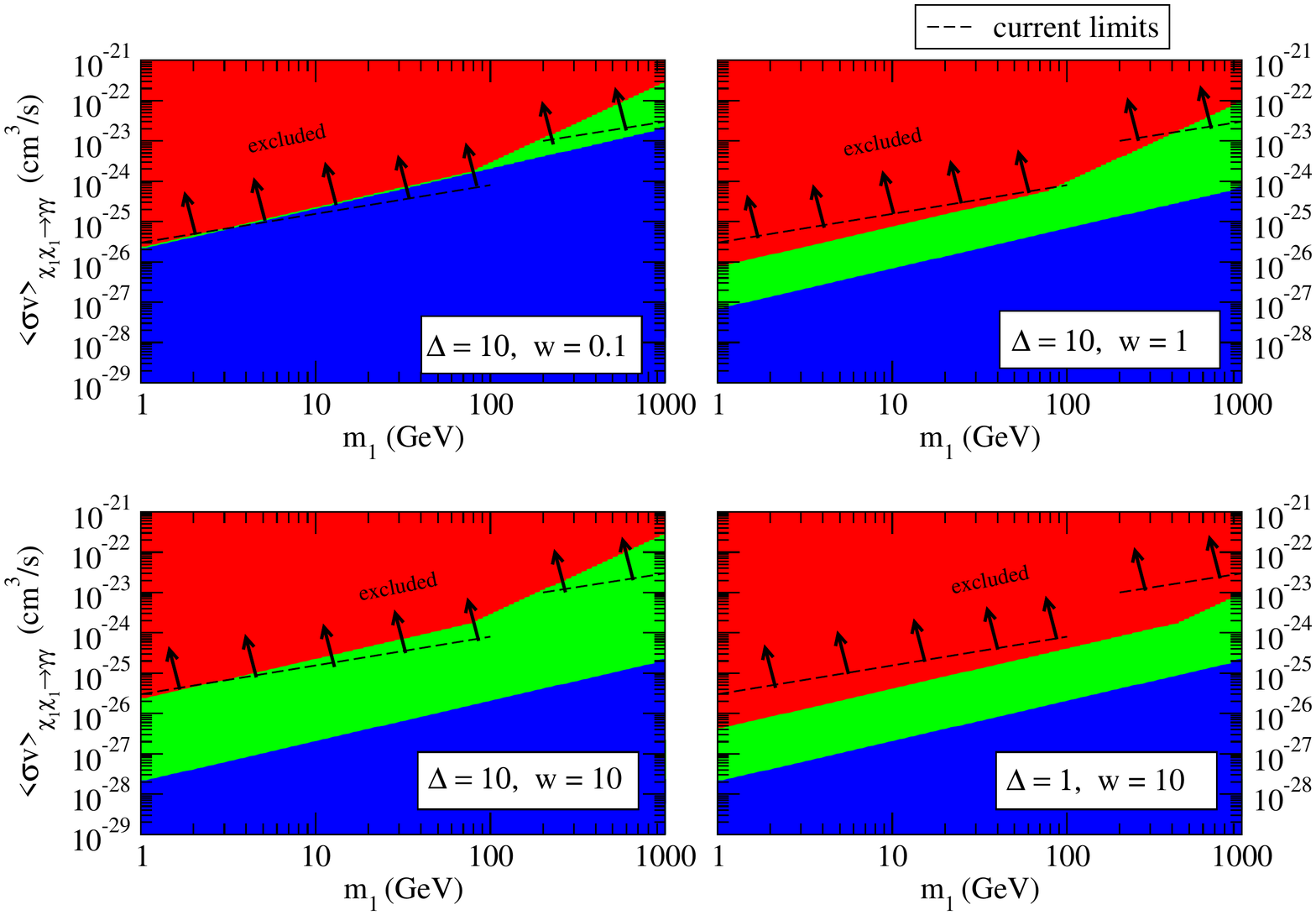}}\\[-1cm]
  \caption{\textit{Indirect detection.} $\langle\sigma_{\chi_{1}\chi_{1}\to\gamma\gamma}v\rangle$ vs $m_1$ is plotted at fixed values of $\Delta$ and $w$, as indicated in each panel. The current limits on gamma-ray lines are taken from the analysis of Ref.~\cite{Mack:2008wu}. The arrows indicate the excluded region. {\em Color coding}: blue stands for {\em no detection}, green for {\em detection without discrimination}, red for {\em detection with discrimination}.}
  \label{fig:iS1m1}
\end{figure}

Fig.~\ref{fig:iS1m1} studies the $(m_1,\langle\sigma_{\chi_{1}\chi_{1}\to\gamma\gamma}v\rangle)$ plane for fixed values of the mass splitting $\Delta$ and of the relative abundance of the two WIMPs, $w$. The dashed lines indicate the current constraints on monochromatic gamma-ray lines as obtained in the analysis of Ref.~\cite{Mack:2008wu} (see also \cite{Pullen:2006sy}), and the regions above those lines are excluded. We are interested in looking for setups where the current exclusion limits fall well above the line that distinguish discrimination from detection, i.e. the boundary between the red and the green regions.

First, we notice that in general, and not unexpectedly, the most favorable scenarios are those where $w\sim1$, i.e. where the dark matter is roughly half made up of the first WIMP species $\chi_1$ and half of the second species $\chi_2$ (e.g. top three panels to the right). When the dark matter is mostly made up of one WIMP species, be it $\chi_2$ (small values of $w\sim0.1$) or $\chi_1$ (large values of $w\sim10$), discrimination is almost never possible. The one exception we find is the panel in the lower right corner, with $\Delta=1$ and $w=10$, where discrimination would be possible with a pair annihilation rate within an order of magnitude of the current limits. Discrimination is also marginally possible for $w=0.1$ and again $\Delta=1$ (second panel from the top in the left column). We thus conclude that for indirect detection, the ideal setup for the relative mass splitting and abundance is when they both are close to 1, i.e. when the two WIMP species are relatively close in mass, and they contribute roughly equally to the dark matter content of the Universe.

\begin{figure}[!ht]
  \centering  
  \includegraphics[width=172mm]{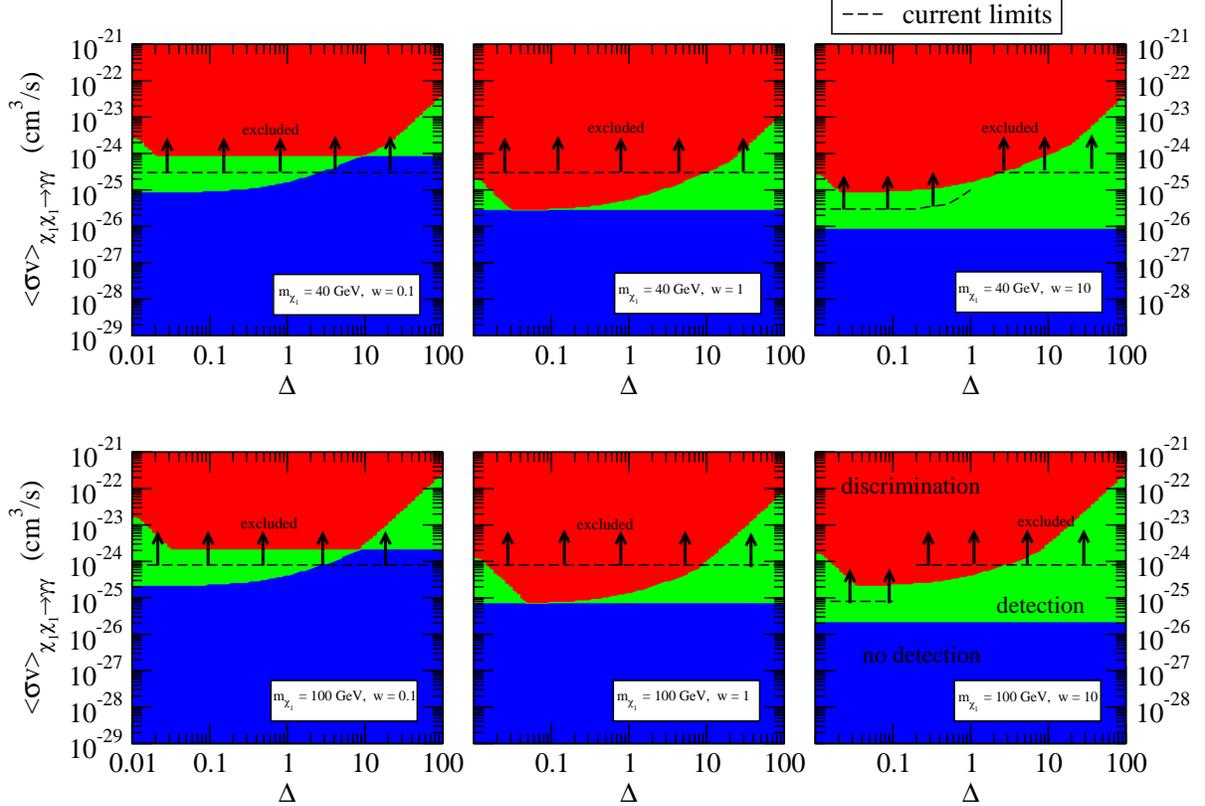}\\[-1cm]
  \caption{\textit{Indirect detection.} $\langle\sigma_{\chi_{1}\chi_{1}\to\gamma\gamma}v\rangle$ vs $\Delta$ is plotted at fixed values of $m_1$ and $w$. The current limits shown are taken from \cite{Mack:2008wu}. The arrows indicate the excluded region. {\em Color coding}: blue stands for {\em no detection}, green for {\em detection without discrimination}, red for {\em detection with discrimination}.} 
  \label{fig:iS1Delta}
\end{figure}

In Fig.~\ref{fig:iS1Delta} we see again that the best chances for discrimination of a dual-component dark matter scenario correspond to $w=1$. In that case, the range of values of $\Delta$ for which discrimination is possible is rather wide, $0.01\lesssim\Delta\lesssim10$, but is optimal for $\Delta$ between 0.1 and 1. As we found before, discrimination is also possible for $w=10$ if $\Delta\sim1$ (see the bottom panel to the right in particular).

\begin{figure}[!ht]
  \centering
  \includegraphics[width=172mm]{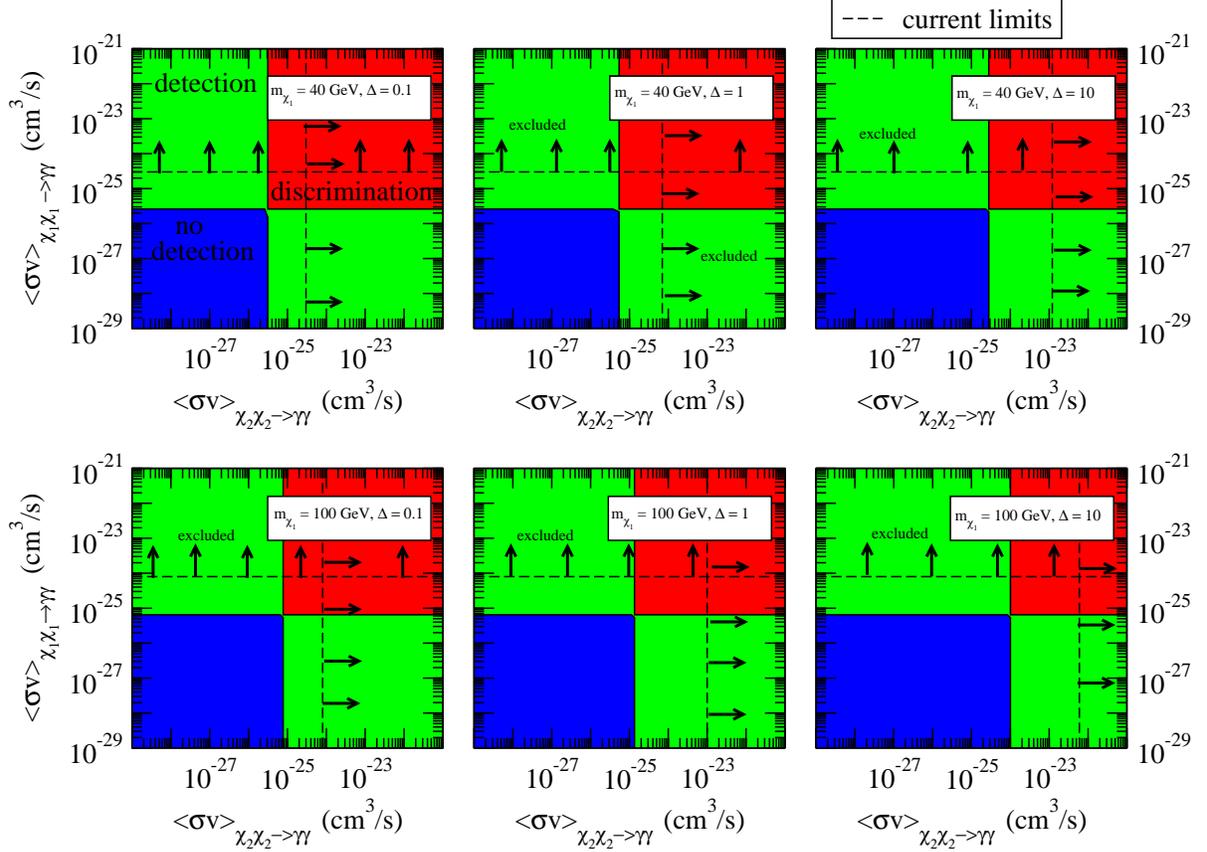}\\[-1cm]
  \caption{\textit{Indirect detection.} Here we relax the assumption (\ref{eq:wsigmas}) and we plot $\langle\sigma_{\chi_{1}\chi_{1}\to\gamma\gamma}v\rangle$ vs $\langle\sigma_{\chi_{2}\chi_{2}\to\gamma\gamma}v\rangle$ at fixed values of $m_1$ and $\Delta$ for equal energy densities $\Omega_{\chi_1} = \Omega_{\chi_2}$ or, equivalently, for $w=1$. The current limits shown are taken from \cite{Mack:2008wu}. The arrows indicate the excluded region. {\em Color coding}: blue stands for {\em no detection}, green for {\em detection without discrimination}, red for {\em detection with discrimination}.} 
  \label{fig:iS1S2}
\end{figure}
 
In Fig.~\ref{fig:iS1S2} we relax the assumption that $\langle\sigma_{\chi_{2}\chi_{2}\to\gamma\gamma}v\rangle=w\times \langle\sigma_{\chi_{1}\chi_{1}\to\gamma\gamma}v\rangle$, and study the ($\langle\sigma_{\chi_{1}\chi_{1}\to\gamma\gamma}v\rangle,\times \langle\sigma_{\chi_{2}\chi_{2}\to\gamma\gamma}v\rangle$) plane. We fix $m_1=40$ GeV in the upper panels and 100 GeV in the lower panels, and set $\Delta=0.1,$ 1 and 10 from left to right. In all cases, we find that there is a range of values for the cross sections where discrimination is possible and consistent with current constraints. This range typically involves cross sections of at least a few$\times 10^{-26}\ {\rm cm}^3/{\rm s}$ ballpark, and extends up to rather large values (particularly when one of the two species is subdominant in the overall dark matter density). We therefore find that for WIMPs with total pair annihilation cross sections close to the typical values needed to achieve thermal production in the early Universe (i.e. $\langle\sigma v\rangle\sim3\times 10^{-26}$) the branching ratio into the $\gamma\gamma$ channel must be extremely large, if not the dominant one. 

\begin{figure}[!ht]
  \centering
  \subfigure{
    \includegraphics[width=150mm]{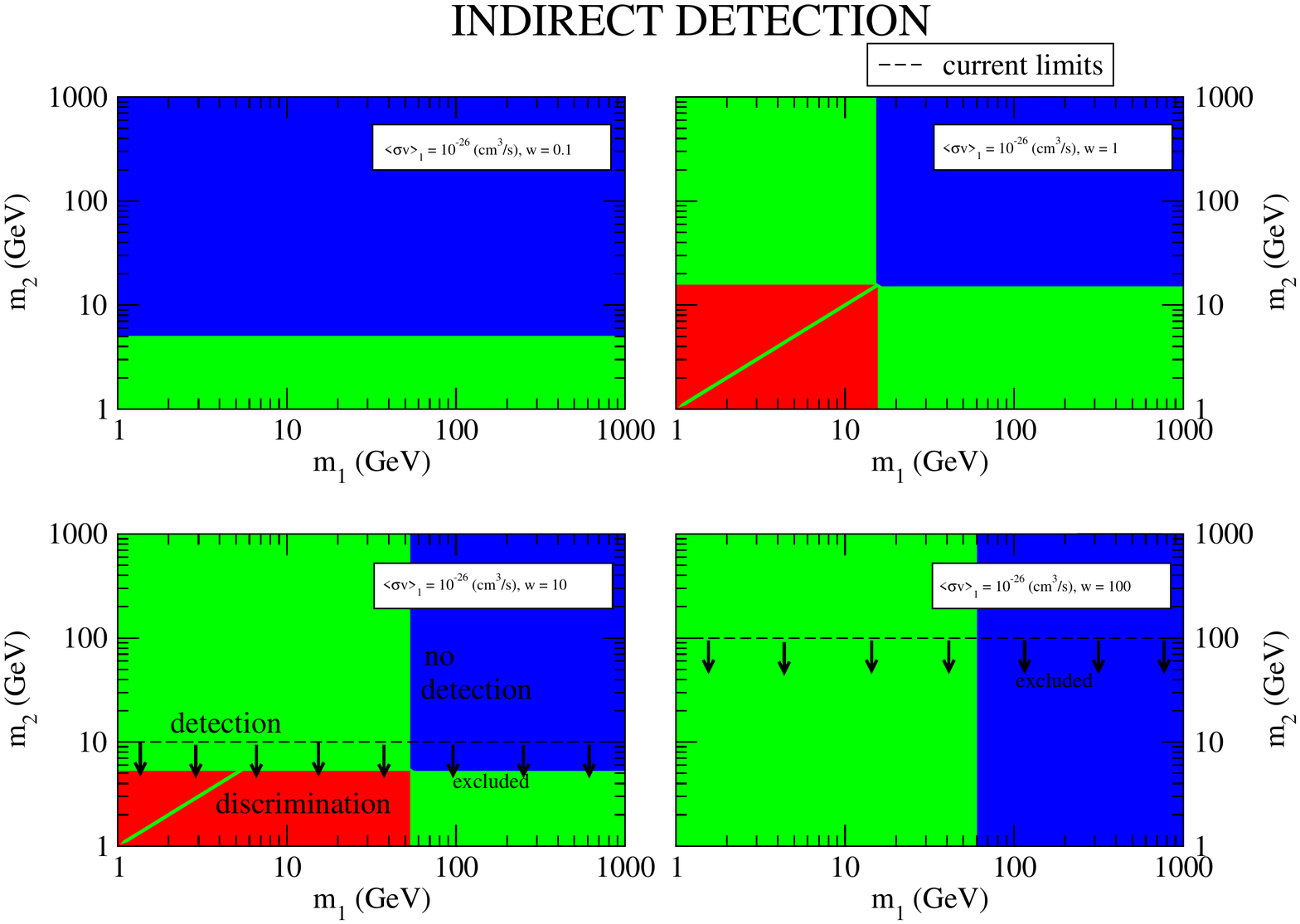}}\\[-1cm]
  \subfigure{
    \includegraphics[width=150mm]{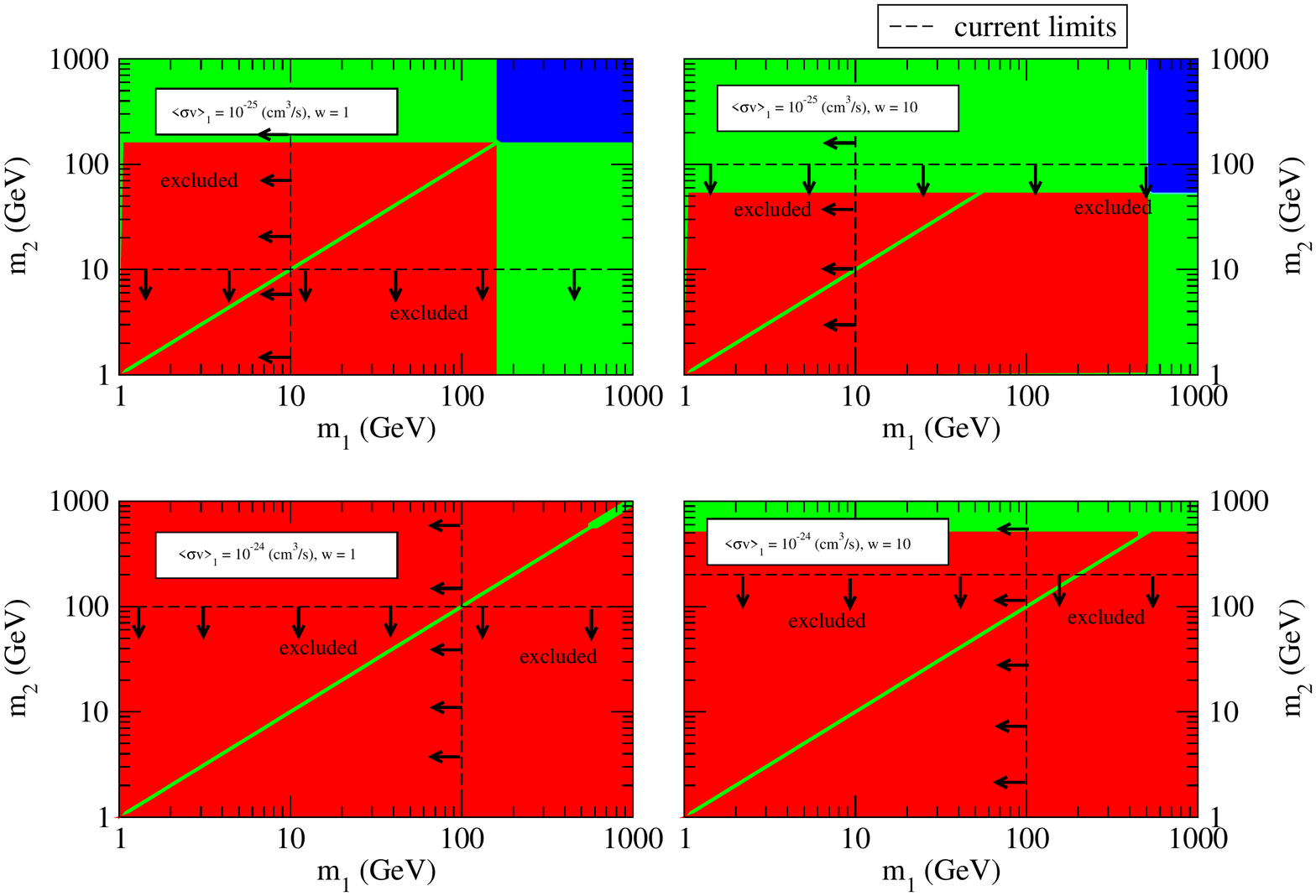}}\\[-1cm]
  \caption{\textit{Indirect detection.} Here we replace $\Delta$ with $m_2$ in the parameter space (\ref{eq:indpar}) and plot $m_2$ vs $m_1$ at fixed values of $\langle\sigma_{\chi_{1}\chi_{1}\to\gamma\gamma}v\rangle$ and $w$. The current limits shown are taken from \cite{Mack:2008wu}. The arrows indicate the excluded region. {\em Color coding}: blue stands for {\em no detection}, green for {\em detection without discrimination}, red for {\em detection with discrimination}.}
  \label{fig:im2m1}
\end{figure}

In Fig.~\ref{fig:im2m1}, we study the plane defined by the two WIMP masses (we relax here the assumption that $m_2>m_1$), and assume a fixed pair annihilation cross section $\langle\sigma_{\chi_{1}\chi_{1}\to\gamma\gamma}v\rangle=10^{-26}\ {\rm cm^3}/{\rm s}$. We then simply consider four different values of $w$, a quantity which also sets the pair annihilation rate into two photons of the second WIMP. While in the top left panel the dark matter is almost entirely made up of $\chi_2$, and that's the only species that can be detected in the channel we consider here (in the bottom right the same is true for $\chi_1$), the most promising cases are again the intermediate ones. For the particular choice of parameters we made, $w=10$ does not lead to any viable discrimination parameter space which is not already excluded by current data, while the ``symmetric'' $w=1$ case allows for discrimination as long as both WIMPs are (1) not degenerate (green line at $m_1=m_2$, where clearly no discrimination is possible) and (2) both lighter than roughly 15 GeV. This precise mass range obviously depends upon the assumed value for $\langle\sigma_{\chi_{1}\chi_{1}\to\gamma\gamma}v\rangle$. 

This point is shown explicitly in the bottom four panels, where we set  $\langle\sigma_{\chi_{1}\chi_{1}\to\gamma\gamma}v\rangle=10^{-25}\ {\rm cm^3}/{\rm s}$ (top) and  $\langle\sigma_{\chi_{1}\chi_{1}\to\gamma\gamma}v\rangle=10^{-24}\ {\rm cm^3}/{\rm s}$ (bottom). Again, we find that the ideal situation is a symmetric case ($w=1$), with masses that can be detectable up to 150 GeV for a cross section of $10^{-25}\ {\rm cm^3}/{\rm s}$ and above a TeV for $10^{-24}\ {\rm cm^3}/{\rm s}$. With $w=10$ discrimination is possible only for large masses and for very large pair annihilation cross sections into two photons ($10^{-24}\ {\rm cm^3}/{\rm s}$, bottom right panel).

\section{Related Work and Potential  Models}\label{sec:related}

The possibility of a ``complex'', multi-partite dark matter sector has been envisioned in different contexts, including models where the dark matter is partly made of WIMPs and partly made of e.g. axions \cite{withaxions} or sterile neutrinos \cite{warmcold}. In addition, models with {\em multiple WIMP} species as dark matter candidates have been discussed, motivated either by theoretical considerations (see e.g. \cite{Ma:2006uv,Hur:2007ur,Cao:2007fy,Feng:2008ya}), or to explain anomalous signatures in cosmic ray or direct detection experiments that could not be easily accounted for with a single dark matter particle (e.g. \cite{Boehm:2003ha,Fairbairn:2008fb,Zurek:2008qg}).

Without claiming to be exhaustive in our account, we summarize here a few recent studies that addressed, in specific particle dark matter models, the possibility that the dark matter sector is comprised of more than one, stable weakly interacting massive particle (we therefore do not count here as ``multi-partite'' (WIMP) dark matter models those frameworks where e.g. besides a stable WIMP one has a contribution to the dark matter density from neutrinos (for instance in mixed warm-cold dark matter models with sterile neutrinos \cite{warmcold}) or from axions \cite{withaxions}). 

Ref.~\cite{Boehm:2003ha} showed that a two-component dark matter scenario (invoked in their analysis to explain both the Integral/SPI 511 keV line in the galactic bulge and the EGRET GeV excess, the first anomaly requiring a light dark matter WIMP in the MeV range and the second one in the GeV range) can emerge in the context of $N=2$ supersymmetry with two discrete symmetries, the ``usual'' $R$ parity of the MSSM and an additional $M$-parity. The lightest $R-$ and $M-$ parity odd particles are stable, the latter in principle have desirable features to fit the Integral/SPI 511 keV line.

In Ref.~\cite{Ma:2006uv}, a scenario was proposed for the radiative generation of Majorana neutrino masses via the introduction of additional superfields to the MSSM, namely three heavy neutral Majorana fermion singlets, two more SU(2) Higgs doublets (in addition to the usual two MSSM Higgs doublets), one scalar singlet. These superfields are odd under a new discrete parity $Z_2^\prime$. The model not only generates naturally small neutrino masses, but it can also account for the baryon asymmetry via leptogenesis, and provides a multi-partite dark matter scenario involving the lightest odd-charged particles under the two discrete symmetries. The phenomenology of this scenario, as well as of the more general setup where the dark matter is comprised of a fermion singlet and of a scalar singlet, with quartic interactions with the SU(2) Standard Model Higgs, was studied in detail in Ref.~\cite{Cao:2007fy}. That study makes several specific and model-dependent assumptions on the
particle physics model, and uses a different parameterization and
definition of whether a model is or not detectable. A direct comparison of
the results presented here with those in Ref.~\cite{Cao:2007fy} is thus arduous,
although Ref.~\cite{Cao:2007fy} presents a specific model realization of the setup we
hereby investigated.

A scenario that incorporates a solution to the problem of small neutrino masses, low-scale leptogenesis and a multi-partite dark matter scenario, involving a $Z_2\times Z_2^\prime$ discrete symmetry, was also recently envisioned in Ref.~\cite{SungCheon:2008ts}. The two dark matter particles of that scenario are a gauge-singlet Majorana neutrino $S$ and a singlet scalar $\phi$. In addition to the phenomenology of the model at direct dark matter detection experiments and with colliders (specifically, through the search for invisible Higgs decays at the LHC), Ref.~\cite{SungCheon:2008ts} also discusses the possibility -- which we do not entertain in the present analysis -- of pair-annihilation processes of the type $SS\to\phi\phi$ or, depending on the mass spectrum, $\phi\phi\to SS$. Those processes can play an important role in setting the relic abundance of the two dark matter particle species in the early universe \cite{SungCheon:2008ts}.

An interesting realization of a two-component dark matter model was presented in Ref.~\cite{Hur:2007ur}, which focused on an extension of the MSSM featuring a spontaneously broken $U(1)^\prime$ gauge symmetry giving rise to a conserved residual discrete $U$-parity, and to an additional stable WIMP that can contribute to the dark matter in addition to the ordinary LSP. This possibility is soundly motivated, e.g. from possible solutions to the MSSM $\mu$ problem and to the suppression of higher dimensional $R$-parity conserving operators that can mediate lepton and baryon number violating interactions. Ref.~\cite{Hur:2007ur} considers the phenomenology of direct detection and collider searches for both the case where the scalar or the fermion component of the $U$-parity odd superfield is the additional lightest stable particle.

Ref.~\cite{Feng:2008ya} pointed out that in gauge mediated supersymmetry breaking it is natural to expect multiple stable hidden sector particles that quite remarkably have the desired thermal relic abundance to explain the dark matter density in the Universe. Along similar lines, this possibility was further investigate in Ref.~\cite{Morrissey:2009ur}. The recent lepton anomalies reported by the Pamela \cite{Adriani:2008zr} and ATIC \cite{:2008zzr} experiments (the latter only seen to a lesser extent by the Fermi gamma-ray space telescope, \cite{Abdo:2009zk,Grasso:2009ma}) have triggered model-building that included the possibility of a multi-partite dark matter setup. This includes e.g. the study of Ref.~\cite{Fairbairn:2008fb}, where the large boost factors needed for the annihilation rate of dark matter particles needed to explain the lepton anomalies are made compatible with a large enough relic abundance via the decays of a meta-stable heavier particle species, that is the dominant dark matter component prior to its decay. Ref.~\cite{Zurek:2008qg} studied instead the possibility that several anomalous signals that might be explained with two particle dark matter particles: one is a heavy quasi-sterile neutrino, pair annihilating preferentially into leptons, as needed to explain the mentioned lepton anomalies, and the second one is a possibly lighter particle that could account for the DAMA signal \cite{Bernabei:2008yi,Fairbairn:2008fb,Zurek:2008qg}.

Unlike in the work summarized above, we have taken in this study an entirely model-independent approach, and do not refer to any specific multi-partite particle dark matter setup. To a good approximation our findings can be appropriately applied and translated to any of these particular models. 

\section{Conclusions}\label{sec:conclusions}

We have focused in this study on the prospects for establishing, with dark matter searches and a future linear collider, the existence of dark-matter physics featuring two stable weakly interacting massive particles in our Universe. Here we summarize our general results, and outline, under our assumptions, the generic requirements on the parameter space of this scenario in order for significant experimental signatures to be imprinted upon one of the channels under consideration.
\begin{itemize}

\item {\em Direct Detection}: we considered the possibility of discerning a multi-component dark matter scenario via fits to recoil spectra in direct detection experiments with a setup of 5 yr of data taking with a 1000 kg target. We found that:
\begin{itemize}
\item To have discrimination of two dark matter species at least one of the WIMPs must have a light mass, in the 10 GeV to 30 GeV range, even with large detectors. The mass of the heavier species can be as large as a TeV, for sufficiently large experimental setups and scattering cross sections.
\item A necessary condition for the discovery of multi-partite dark matter with direct detection experiments is that the two species contribute comparably to the overall dark matter density of the Universe.
\item The relative mass splitting $\Delta$ for which we find that it would be possible to discover multi-component dark matter is rather narrow for direct detection, ranging between 0.5 and 2, and peaking where the mass ratio between the two dark matter species is in the proportion of 2:1 ($\Delta=1$).
\item We do not find the choice of the nuclear target to be crucial to discover a multi-component dark matter scenario.
\end{itemize}

\item {\em Linear Collider}: we assumed that in a future electron-positron linear collider with a center of mass energy of 0.5 TeV a pair of charged massive particles is produced and decays into one of the two stable WIMPs plus a light lepton, with a branching fraction related to the relative WIMP abundance in the overall dark matter density. We found that:
\begin{itemize}
\item When the decay of the massive charged species into the WIMPs is kinematically open for both, the discovery of two stable species (on collider scales) is possible for relative mass splitting $\Delta$ as small as 0.05 and up to 2.
\item The best case scenario is again when the relative abundance of the two species is comparable, which, with our assumptions, corresponds to comparable branching ratio into the two WIMPs for the decay modes of the charged massive species.
\item Discovery prospects worsen near the kinematic limits of the collider, simply due to lack of statistics, but the reach extends almost all the way to the kinematic limit in most cases.
\end{itemize}

\item {\em Indirect Detection:} we considered the detectability of two distinct gamma-ray lines associated with the direct annihilation of the WIMPs into monochromatic gamma rays. Because of the inability to distinguish whether the second line comes from a second WIMP or from a different decay mode of the first WIMP, this search channel cannot lead to a discovery in its own right. Nevertheless, it could prove useful for a verification of the dual-component dark matter scenario.  We found that:
\begin{itemize}
\item Large pair annihilation cross sections in the $\gamma\gamma$ channel are needed to have a detectable signature of two stable WIMPs, typically of the order of $10^{-26}\ {\rm cm}^3/s$, or at most two orders of magnitude below the current, pre-Fermi-LAT experimental sensitivity.
\item Under the assumption (which could be relaxed) that the total annihilation cross section is proportional to the annihilation cross section into $\gamma\gamma$, this detection channel strongly depends upon the relative composition of the dark matter; the possibility to find two stable WIMP species is strongly favored if the two species contribute almost equally to the overall dark matter density.
\item Indirect detection of dual-component dark matter is quite effective in a wide range of masses, and for a considerably large range of relative mass splittings between the two species' masses, $\Delta$, which can range between 0.01 up to more than 10.
\end{itemize}

\end{itemize}

\section*{Acknowledgements}

We would like to thank Bruce Schumm and Michael Dine for useful discussions.  K.S. thanks
Aspen Center for Physics, where part of this work was completed, for its hospitality.
S.P. is partly supported by an Outstanding Junior Investigator Award from the US Department of Energy (DoE), Office of Science, High Energy Physics, and by DoE Contract DEFG02-04ER41268 and NSF Grant PHY-0757911. S.P. and L.U. are partly supported by NASA Grant Number NNX08AV72G. K.S. is supported by a Natural Sciences and Engineering Research Council (NSERC) of Canada Discovery Grant.

\end{document}